\documentclass{jfm}
\usepackage{graphicx}
\usepackage{tikz,xcolor}
\usepackage{graphicx,subfigure, array}
\usepackage{natbib}
\usepackage{multirow}
\usepackage{amsmath}
\usepackage{amsfonts,amssymb}
\usepackage[switch, modulo]{lineno}
\usepackage[colorlinks=true,citecolor = blue]{hyperref}

\usepackage{tikz,xcolor,hyperref}

\definecolor{lime}{HTML}{A6CE39}
\DeclareRobustCommand{\orcidicon}{%
	\begin{tikzpicture}
	\draw[lime, fill=lime] (0,0) 
	circle [radius=0.16] 
	node[white] {{\fontfamily{qag}\selectfont \tiny ID}};
	\draw[white, fill=white] (-0.0625,0.095) 
	circle [radius=0.007];
	\end{tikzpicture}
	\hspace{-2mm}
}

\foreach \x in {A, ..., Z}{%
	\expandafter\xdef\csname orcid\x\endcsname{\noexpand\href{https://orcid.org/\csname orcidauthor\x\endcsname}{\noexpand\orcidicon}}
}

\shorttitle{Precession-driven flows in stress-free ellipsoids}
\shortauthor{J. Vidal \& D. C\'ebron}

\title{Precession-driven flows in stress-free ellipsoids}

\author{J\'er\'emie Vidal
  \corresp{\email{jeremie.vidal@univ-grenoble-alpes.fr}} \& David C\'ebron}

\affiliation{Universit\'e Grenoble Alpes, CNRS, ISTerre, 38000 Grenoble, France}

\begin{document}

\maketitle

\begin{abstract}
Motivated by modelling rotating turbulence in planetary fluid layers, we investigate precession-driven flows in ellipsoids subject to stress-free boundary conditions (SF-BC). 
The SF-BC could indeed unlock numerical constraints associated with the no-slip boundary conditions (NS-BC), but are also relevant for some astrophysical applications. 
Although SF-BC have been employed in the pioneering work of Lorenzani \& Tilgner (\emph{J. Fluid Mech.}, 2003, \textbf{492}, pp. 363--379), they have scarcely been used due to the discovery of some specific mathematical issues associated with angular momentum conservation. 
We revisit the problem using asymptotic analysis in the low-viscosity regime, which is validated with numerical simulations.
First, we extend the reduced model of uniform-vorticity flows in ellipsoids to account for SF-BC.
We show that the long-term evolution of angular momentum is affected by viscosity in triaxial geometries, but also in axisymmetric ellipsoids when the mean rotation axis of the fluid is not the symmetry axis.
In a regime relevant to planets, we analytically obtain the primary forced flow in triaxial geometries, which exhibits a second inviscid resonance. 
Then, we investigate the bulk instabilities existing in precessing ellipsoids. 
We show that using SF-BC would be useful to explore the non-viscous instabilities (e.g. Kerswell, \emph{Geophys. Astrophys. Fluid Dyn.}, 1993, \textbf{72}, pp. 107-144),
which are presumably relevant for planetary applications but are often hampered in experiments or simulations with NS-BC.
\end{abstract}


\section{Introduction}
Motivated by numerous natural applications \citep[e.g.][]{le2015flows}, we aim to explore the long-term dynamics of rapidly rotating fluids enclosed in ellipsoids subject to (harmonic) mechanical forcings. 
Global rotation is indeed ubiquitous in many planetary fluid layers or stars, which are usually ellipsoidal at the leading order \citep[e.g. due to the combined action of centrifugal effects and gravitational interactions with nearby orbital partners, see][]{chandrasekhar1969ellipsoidal}. 
In particular, mechanically driven flows in ellipsoids (e.g. flows driven by precession or tides) have received much attention in the fluid community.
Mechanical forcings can indeed sustain bulk instabilities \citep[e.g.][]{kerswell1993instability,kerswell2002elliptical}, turbulence \citep[e.g.][]{grannan2017tidally,le2019experimental} and possibly dynamo magnetic fields \citep[e.g.][]{reddy2018turbulent,vidal2018magnetic}. 
These works have also renewed interest in a key fundamental question in the theory of rotating fluids, which is the generation of two-dimensional geostrophic motions \citep{greenspan1969non}.
However, this problem has only received scant attention in global geometries exhibiting the so-called topographic beta effect \citep[which strongly modifies the geostrophic flows, e.g.][]{greenspan1968theory}. 
Exploring rotating turbulence thus deserves further work using global models. 

The incompressible Navier-Stokes equation is commonly adopted to explore the turbulence driven by mechanical forcings, together with the no-slip boundary conditions (NS-BC).
The latter are appropriate to model the flow dynamics in the presence of a rigid boundary (e.g. the solid interface between a liquid core and a solid overlying mantle in planetary interiors). 
However, the range of parameters that is accessible to global simulations with NS-BC is severely limited, in particular for the Ekman number $E$ (which crucially controls the dynamics of rapidly rotating flows). 
Typical values in natural systems are $E \leq \mathcal{O}(10^{-12})$, whereas direct numerical simulations (DNS) and laboratory experiments of mechanically driven rotating turbulence can only reach much larger values $E \gtrsim 10^{-6}$ \citep[e.g.][]{grannan2017tidally,le2019experimental}. 
As a consequence, the Ekman boundary layer is often a prominent feature in the models (whereas the smallness of $E$ in planetary systems suggests that viscosity should rather play a minor dynamical role), and its resolution requires considerable computational resources when $E$ is lowered. 
Moreover, the overestimated viscous torque at the boundary can also largely inhibit the fluid response to mechanical forcings (which is primarily driven by the shape deformation of the fluid boundary, combined with non-stationary effects due to the possibly oscillatory angular velocity of the container). 
Therefore, different modelling approaches are worth considering to simulate such flows at more realistic parameters for planetary applications. 

One natural way to avoid the physical and computational disadvantages of NS-BC is to employ stress-free boundary conditions (SF-BC). 
A thin outer Ekman boundary layer is still present for stress-free boundaries \citep[e.g.][]{livermore2016comparison}, but its dynamical role is expected to be less important because the boundary-layer flow is much weaker in amplitude than the bulk flow \citep[e.g.][]{rieutord1992ekman}. 
Moreover, SF-BC are also commonly employed in astrophysical modelling since they are often believed to yield similar results to those obtained with a realistic free surface \citep{barker2016non}. 
However, SF-BC have scarcely been used in spheres and ellipsoids because of mathematical difficulties. 
The most serious one is related to angular momentum conservation. 
Angular momentum can indeed be arbitrary in axisymmetric geometries, leading to spurious solutions on long time scales \citep[e.g.][]{jones2011anelastic,guermond2013remarks}. 
The usefulness of SF-BC for simulating rotating flows in ellipsoids has thus been questioned, but we believe that this mathematical set-up deserves further analysis.

In this paper, we thus revisit the influence of SF-BC for rotating ellipsoids using asymptotic analysis when $E \ll 1$ and targeted numerical simulations.
The paper is organised as follows. 
The model is presented in \S\ref{sec:formulation} and applied to precessing ellipsoids in \S\ref{sec:precession}.
The results are discussed in \S\ref{sec:discussion}, and we end the paper in \S\ref{sec:conclusion}.

\section{Mathematical modelling}
\label{sec:formulation}
\subsection{Fluid dynamic equations}
We consider a fluid-filled ellipsoid of uniform density and volume $V$, which is assumed to co-rotate with the surrounding mantle at the angular velocity $\boldsymbol{\Omega}_c (t) = \Omega_0 \, [ \boldsymbol{\Omega} + \boldsymbol{\delta}(t) ]$ with respect to the inertial frame ($\boldsymbol{\delta}(t)$ being the time-dependent departure from the steady global rotation $\boldsymbol{\Omega}$ along the unit vector $\boldsymbol{1}_\Omega = \boldsymbol{\Omega}/|\boldsymbol{\Omega}|$). 
To have a tractable mathematical problem, we seek mechanically driven flows in the mantle reference frame in which the ellipsoidal boundary $S$ is steady and $\boldsymbol{\delta}(t) \neq \boldsymbol{0}$. 
This set-up allows us to model flows driven by precession or librations, which have already received consideration using NS-BC \citep[e.g.][]{noir2013precession,zhang2012asymptotic,zhang2014precessing,vantieghem2015latitudinal}. 
We non-dimensionalise the problem using $\Omega_0^{-1}$ as the time scale, and a typical length $R$ as the length scale (which is here arbitrary).
Considering a Newtonian fluid of uniform kinematic viscosity $\nu$, the dimensionless equations for the velocity $\boldsymbol{v}$ are
\begin{subequations}
\allowdisplaybreaks
\label{eq:NSincomp}
\begin{align}
	\partial_t \boldsymbol{v} + ( \boldsymbol{v} \boldsymbol{\cdot} \nabla ) \, \boldsymbol{v} + 2 \boldsymbol{\Omega}_c \times \boldsymbol{v} &= - \nabla p + 2 E \, \nabla \boldsymbol{\cdot} \boldsymbol{\epsilon}(\boldsymbol{v}) + \boldsymbol{r} \times \mathrm{d}_t \boldsymbol{\delta}, \\
	\nabla \boldsymbol{\cdot} \boldsymbol{v} &= 0,
\end{align}
\end{subequations}
where $\boldsymbol{r}$ is the position vector, $\boldsymbol{\epsilon} (\boldsymbol{v}) =(1/2) [ \nabla \boldsymbol{v} + (\nabla \boldsymbol{v})^\top ]$ is the strain-rate tensor, and $E = \nu/(\Omega_0 R^2)$ is the Ekman number.
The ellipsoidal geometry, which is assumed to be steady in the mantle frame, is given by the dimensionless equation
\begin{equation}
    (x/a)^2 + (y/b)^2 + (z/c)^2 = 1
\end{equation} 
where $[a,b,c]$ are the (dimensionless) ellipsoidal semi-axes and $[x,y,z]$ are the Cartesian coordinates. 
In the following, axisymmetric geometries refer to ellipsoids with a revolution symmetry axis (i.e. when either $a=b$, $b=c$ or $a=c$).
Finally, spheroids will refer to the particular axisymmetric geometries for which the revolution symmetry axis is aligned with the rotation axis (with $a=b$ and $\boldsymbol{\Omega} \propto \boldsymbol{1}_z$ in this study).  
We aim to consider the SF-BC given in the mantle frame by
\begin{subequations}
\label{eq:BCSF}
\begin{equation}
    \left . \boldsymbol{v} \boldsymbol{\cdot} \boldsymbol{1}_n \right |_S = 0, \quad \left . [ \boldsymbol{\epsilon} (\boldsymbol{v})  \boldsymbol{\cdot} \boldsymbol{1}_n] \times \boldsymbol{1}_n \right |_S = \boldsymbol{0},
    \tag{\theequation \emph{a,b}}
\end{equation}
\end{subequations}
where $\boldsymbol{1}_n$ is the outward normal unit vector at the boundary, instead of the NS-BC 
\begin{subequations}
\label{eq:BCNS}
\begin{equation}
    \left . \boldsymbol{v} \boldsymbol{\cdot} \boldsymbol{1}_n \right |_S = 0, \quad \left . \boldsymbol{v} \times \boldsymbol{1}_n \right |_S = \boldsymbol{0}.
    \tag{\theequation \emph{a,b}}
\end{equation}
\end{subequations}
It is obvious from SF-BC (\ref{eq:BCSF}) and NS-BC (\ref{eq:BCNS}) that the tangential velocity at the boundary will differ between the two cases (since the flow is allowed to freely slip on the boundary with the SF-BC). 
One may thus wonder in which circumstances the above conditions will lead to similar flows in the bulk (i.e. far from the boundary region). 

A necessary condition is that the mechanical forcings can sustain flows against viscous dissipation for the two BC in the mantle frame. 
This is evidenced by the conservation equation for the volume-averaged kinetic energy $E_k$. 
In a frame where the fluid boundary is steady, it is given by \citep[e.g. equation 5 in][]{wu2009dynamo}
\begin{equation}
    \mathrm{d}_t E_k = \int_V \boldsymbol{v} \boldsymbol{\cdot} \left [ \boldsymbol{r} \times \mathrm{d}_t \boldsymbol{\delta} \right ]  \mathrm{d} V + 2 E \left ( \int_S \boldsymbol{v} \boldsymbol{\cdot} \boldsymbol{\mathcal{T}} \, \mathrm{d}S - \mathcal{D}_\nu \right )
    \label{eq:KE}
\end{equation}
where $\boldsymbol{\mathcal{T}} = \boldsymbol{\epsilon}(\boldsymbol{v}) \boldsymbol{\cdot} \boldsymbol{1}_n$ is the surface traction and $\mathcal{D}_v \geq 0$ is a volume-averaged viscous dissipation (for both the NS-BC and SF-BC).
For a velocity satisfying the no-penetration condition such that $\boldsymbol{v} = (\boldsymbol{v} \boldsymbol{\cdot} \boldsymbol{1}_n) \, \boldsymbol{1}_n - \boldsymbol{1}_n \times (\boldsymbol{1}_n \times \boldsymbol{v}) = - \boldsymbol{1}_n \times (\boldsymbol{1}_n \times \boldsymbol{v})$, the surface integral can actually be written as
\begin{equation}
    \int_S \boldsymbol{v} \boldsymbol{\cdot} \boldsymbol{\mathcal{T}} \, \mathrm{d}S =  - \int_S \boldsymbol{\mathcal{T}} \boldsymbol{\cdot}  \left [ \boldsymbol{1}_n \times (\boldsymbol{1}_n \times \boldsymbol{v}) \right ] \mathrm{d} S = - \int_S \left [ \boldsymbol{\mathcal{T}} \times \boldsymbol{1}_n \right ] \boldsymbol{\cdot} \left [\boldsymbol{v} \times \boldsymbol{1}_n \right ] \mathrm{d}S
\end{equation}
where we have used a property of the scalar triple product to obtain the last expression.
Thus, the above surface integral exactly vanishes for both SF-BC (\ref{eq:BCSF}) and NS-BC (\ref{eq:BCNS}) in the mantle frame. 
Then, equation (\ref{eq:KE}) shows that we can have $\mathrm{d}_t E_k \geq 0$ for both SF-BC and NS-BC if the mechanical forcings are oscillatory in the mantle frame (i.e. when $\mathrm{d}_t \boldsymbol{\delta} \neq \boldsymbol{0}$). 
Harmonic mechanical forcings, such as precession or librations, can thus sustain flows against viscous dissipation in the mantle frame (even with the SF-BC).  
Note that a very different conclusion is obtained for steady forcings, such as precession viewed in the frame of precession for spheroidal geometries \citep{lorenzani2003inertial,wu2009dynamo}. 
We indeed have $\mathrm{d}_t E_k < 0$ at every time for the SF-BC in the precession frame, whereas precession could sustain non-vanishing flows against viscous dissipation for the NS-BC \citep[since $\left . \boldsymbol{v} \times \boldsymbol{1}_n \right |_S \neq \boldsymbol{0}$ for a no-slip boundary in the precession frame, e.g.][]{cebron2019precessing}.
In the following, we will only investigate the dynamics driven by oscillatory forcings in the mantle frame with SF-BC.

\subsection{Angular momentum}
The angular momentum $\boldsymbol{L} = \int_V \boldsymbol{r} \times \boldsymbol{v} \ \mathrm{d}V$ of the flow plays a central dynamical role for mechanically driven flows in ellipsoids. 
Actually, the Cartesian components of the angular momentum $\boldsymbol{L} = (L_x, L_y, L_z)^\top$ are exactly given for incompressible flows by
\begin{subequations}
\label{eq:LxLyLzproj}
\allowdisplaybreaks
\begin{align}
    L_x &= \int_V \left ( y v_z - z v_y \right ) \mathrm{d} V = \int_V \boldsymbol{v} \boldsymbol{\cdot} \left ( \boldsymbol{1}_x \times \boldsymbol{r} + \nabla \Psi_x \right )   \mathrm{d}V, \\
    L_y &= \int_V \left ( z v_x - x v_z \right ) \mathrm{d} V = \int_V \boldsymbol{v} \boldsymbol{\cdot} \left ( \boldsymbol{1}_y \times \boldsymbol{r} + \nabla \Psi_y \right )  \mathrm{d}V, \\
    L_x &= \int_V \left ( x v_y - y v_x \right ) \mathrm{d} V = \int_V \boldsymbol{v} \boldsymbol{\cdot} \left ( \boldsymbol{1}_z \times \boldsymbol{r} + \nabla \Psi_z \right ) \mathrm{d}V,
\end{align}
\end{subequations}
where $[\Psi_x,\Psi_y,\Psi_z]$ are arbitrary scalar potentials if $\nabla \boldsymbol{\cdot} \boldsymbol{v} = 0$ and if the flow obeys the no-penetration BC in rigid ellipsoids. 
The scalar potentials are thus often discarded to simply express the angular momentum as projections onto the solid-body rotations $\boldsymbol{1}_i \times \boldsymbol{r}$ \citep[e.g.][]{guermond2013remarks}. 
Yet, the solid-body rotations are not admissible flow solutions in non-spherical geometries (even without viscosity), since they do not satisfy the no-penetration condition.

A more appropriate definition of the angular momentum for incompressible flows is thus given in ellipsoids by
\begin{equation}
    \boldsymbol{L} \boldsymbol{\cdot} \boldsymbol{1}_i = \int_V \boldsymbol{e}_i \boldsymbol{\cdot} \boldsymbol{v} \, \mathrm{d}V,
    \label{eq:projectionL}
\end{equation}
where $\{\boldsymbol{e}_i \}_{i\in\{x,y,z\}}$ is the set of  uniform-vorticity (flow) elements defined by 
\begin{subequations}
\label{eq:GP1}
\begin{equation}
    \boldsymbol{e}_i = \boldsymbol{1}_i \times \boldsymbol{r} + \nabla \Psi_i, \quad \nabla \boldsymbol{\cdot} \boldsymbol{e}_i = 0, \quad \left . \boldsymbol{e}_i \boldsymbol{\cdot} \boldsymbol{1}_n \right |_S = 0.
    \tag{\theequation \emph{a--c}}
\end{equation}
\end{subequations}
The scalar functions $\Psi_i$ allow the elements $\boldsymbol{e}_i$ to satisfy the no-penetration condition. 
In ellipsoidal geometries, they are explicitly given by \citep[e.g.][]{noir2013precession}
\begin{subequations}
\label{eq:PsikGP1}
\begin{equation}
    \Psi_x = \frac{c^2-b^2}{b^2+c^2} yz, \quad \Psi_y =  \frac{a^2-c^2}{a^2+c^2} xz, \quad \Psi_z = \frac{b^2-a^2}{a^2 + b^2} xy.
    \tag{\theequation \emph{a--c}}
\end{equation}
\end{subequations}
It is worth noting that definition (\ref{eq:projectionL}) is purely kinematic. 
It thus remains valid in the presence of additional effects, for instance without global rotation or with magnetic effects \citep[e.g][]{gerick2020pressure}.
Moreover, this definition can also be generalised for compressible flows under the anelastic approximation (see Appendix \ref{appendix:Lcompressible}).  
Consequently, we can always rigorously expand incompressible velocity fields in ellipsoids as
\begin{subequations}
\label{eq:expansionBuffett}
\begin{equation}
    \boldsymbol{v}(\boldsymbol{r},t) = \boldsymbol{U}(\boldsymbol{r},t) + \boldsymbol{v}_f (\boldsymbol{r},t), \quad \int_V \boldsymbol{r} \times \boldsymbol{v}_f \, \mathrm{d} V = \boldsymbol{0},
    \tag{\theequation \emph{a,b}}
\end{equation}
\end{subequations}
where the uniform-vorticity flow $\boldsymbol{U}$ carrying the angular momentum is given by
\begin{subequations}
\label{eq:uGP1}
\begin{equation}
    \boldsymbol{U}(\boldsymbol{r},t) = \omega_x (t) \, \boldsymbol{e}_x (\boldsymbol{r}) + \omega_y (t) \, \boldsymbol{e}_y (\boldsymbol{r}) + \omega_z (t) \, \boldsymbol{e}_z (\boldsymbol{r}), \quad \left. \boldsymbol{U} \boldsymbol{\cdot} \boldsymbol{1}_n \right |_S = 0,
    \tag{\theequation \emph{a,b}}
\end{equation}
\end{subequations}
and with the effective rotation vector of the fluid $\boldsymbol{\omega} (t) =(\omega_x (t), \omega_y (t), \omega_z (t))^\top$. 
The velocity $\boldsymbol{v}_f$, which does not carry angular momentum by definition since $\int_V \boldsymbol{U} \boldsymbol{\cdot} \boldsymbol{v}_f \, \mathrm{d} V~=~0$, contains bulks flows of higher spatial complexity (e.g. flow instabilities or turbulence) and also viscous structures \citep[e.g. the Ekman boundary layer,][]{rieutord1992ekman}. 
The Cartesian components of $\boldsymbol{L}$ are then exactly given by
\begin{subequations}
\label{eq:Linv}
\begin{equation}
    \boldsymbol{L} =\mathcal{L}^{-1} \, \boldsymbol{\omega}, \quad     \mathcal{L}^{-1} = \frac{16\pi}{15} abc \
    \mathrm{diag} \left [ \dfrac{b^2c^2}{b^2+c^2}, \dfrac{a^2c^2}{a^2+c^2}, \dfrac{a^2b^2}{a^2+b^2} \right ].
    \tag{\theequation \emph{a,b}}
\end{equation}
\end{subequations}
Finally, the time evolution of the angular momentum (or equivalently that of $\boldsymbol{\omega}$) is affected by viscosity through the action of the viscous torque $\boldsymbol{\Gamma}_\nu$ on long time scales. 
We have for example $\boldsymbol{\Gamma}_\nu = \boldsymbol{0}$ in spheres, such that angular momentum has to be conserved for uniformly rotating fluids in the inertial frame \citep[e.g.][]{jones2011anelastic}. 
To clarify the dynamical role of SF-BC in ellipsoids, it is worth computing the viscous torque.

\subsection{Viscous torque in stress-free ellipsoids}
\label{subsec:viscoustorque}
Because of definition (\ref{eq:projectionL}), the Cartesian components of the viscous torque $\boldsymbol{\Gamma}_\nu = (\boldsymbol{\Gamma}_{\nu} \boldsymbol{\cdot} \boldsymbol{1}_x, \boldsymbol{\Gamma}_{\nu} \boldsymbol{\cdot} \boldsymbol{1}_y, \boldsymbol{\Gamma}_{\nu} \boldsymbol{\cdot} \boldsymbol{1}_z)^\top$ are exactly given for SF-BC (\ref{eq:BCSF}) by
\begin{equation}
    \boldsymbol{\Gamma}_{\nu} \boldsymbol{\cdot} \boldsymbol{1}_i = 2 E \int_V \boldsymbol{e}_i \boldsymbol{\cdot} \nabla \boldsymbol{\cdot} \boldsymbol{\epsilon}(\boldsymbol{v}) \, \mathrm{d} V = - 2E \int_V \boldsymbol{\epsilon}(\boldsymbol{e}_i) : \boldsymbol{\epsilon}(\boldsymbol{v}) \, \mathrm{d} V,
   \label{eq:viscoustorque1}
\end{equation}
where we have used integration by parts and the decomposition
$\boldsymbol{e}_i = (\boldsymbol{1}_n \boldsymbol{\cdot} \boldsymbol{e}_i) \,  \boldsymbol{1}_n - \boldsymbol{1}_n \times (\boldsymbol{1}_n \times \boldsymbol{e}_i) = - \boldsymbol{1}_n \times (\boldsymbol{1}_n \times \boldsymbol{e}_i)$ to cancel out the surface integral for SF-BC \citep[e.g. see the proof of proposition 2.1 in][]{guermond2013remarks}. 
We recover from the formula that $\boldsymbol{\Gamma}_\nu = \boldsymbol{0}$ in spheres since $\boldsymbol{\epsilon}(\boldsymbol{e}_i)$ exactly vanishes when $\boldsymbol{e}_i$ is a solid-body rotation, but we also obtain that  $\boldsymbol{\Gamma}_\nu \neq \boldsymbol{0}$ in triaxial geometries (because $\boldsymbol{\epsilon}(\boldsymbol{e}_i) \neq 0$ when $a\neq b \neq c$).
Moreover, it shows that $\boldsymbol{\Gamma}_{\nu} \boldsymbol{\cdot} \boldsymbol{1}_i =0$ when the Cartesian vector $\boldsymbol{1}_i$ is an axis of revolution of the geometry (irrespective of the fluid global rotation, as $\boldsymbol{e}_i$ is then a solid-body rotation).

\begin{figure}
    \centering
    \begin{tabular}{cc}
    \includegraphics[width=0.49\textwidth]{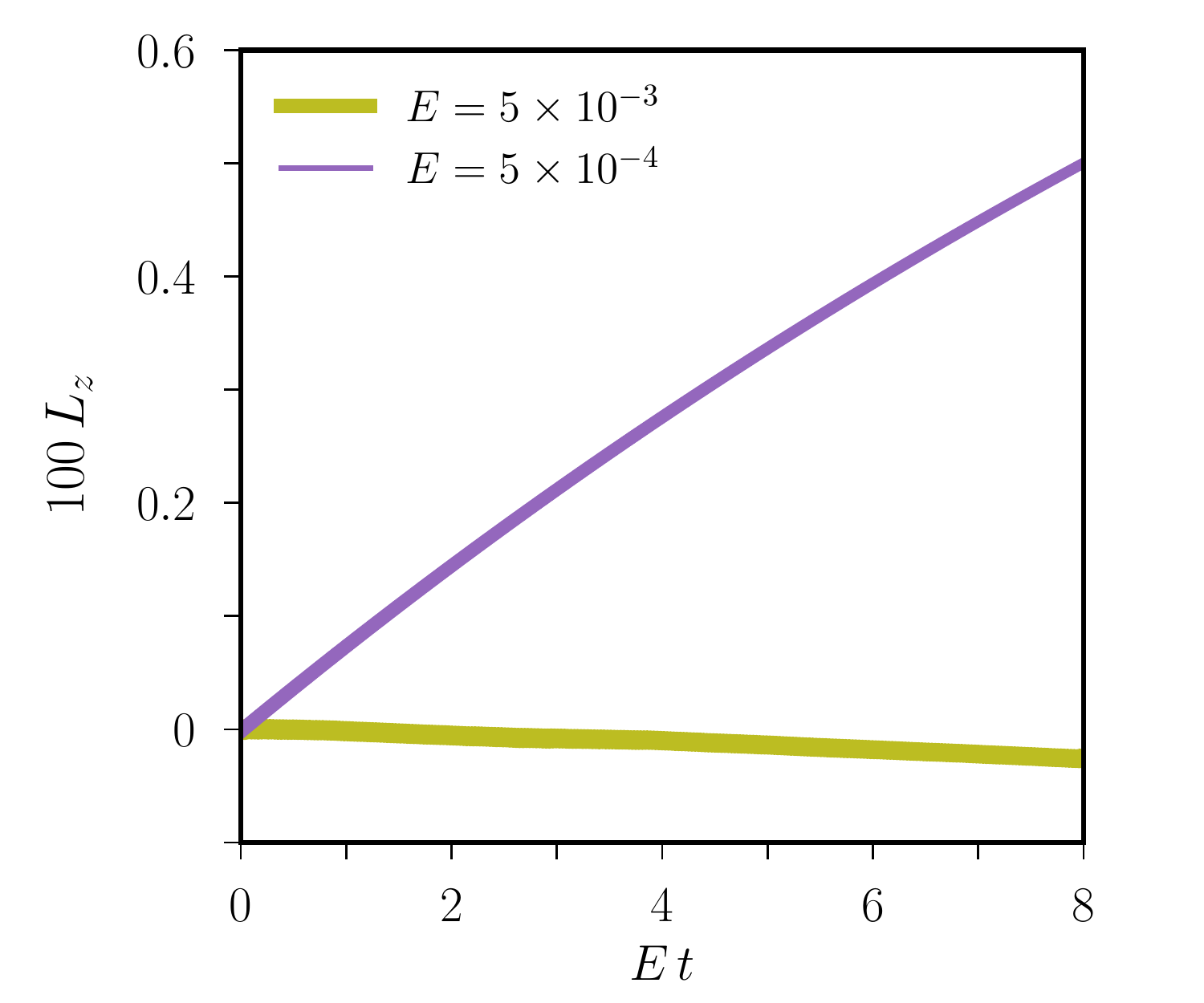} & 
    \includegraphics[width=0.49\textwidth]{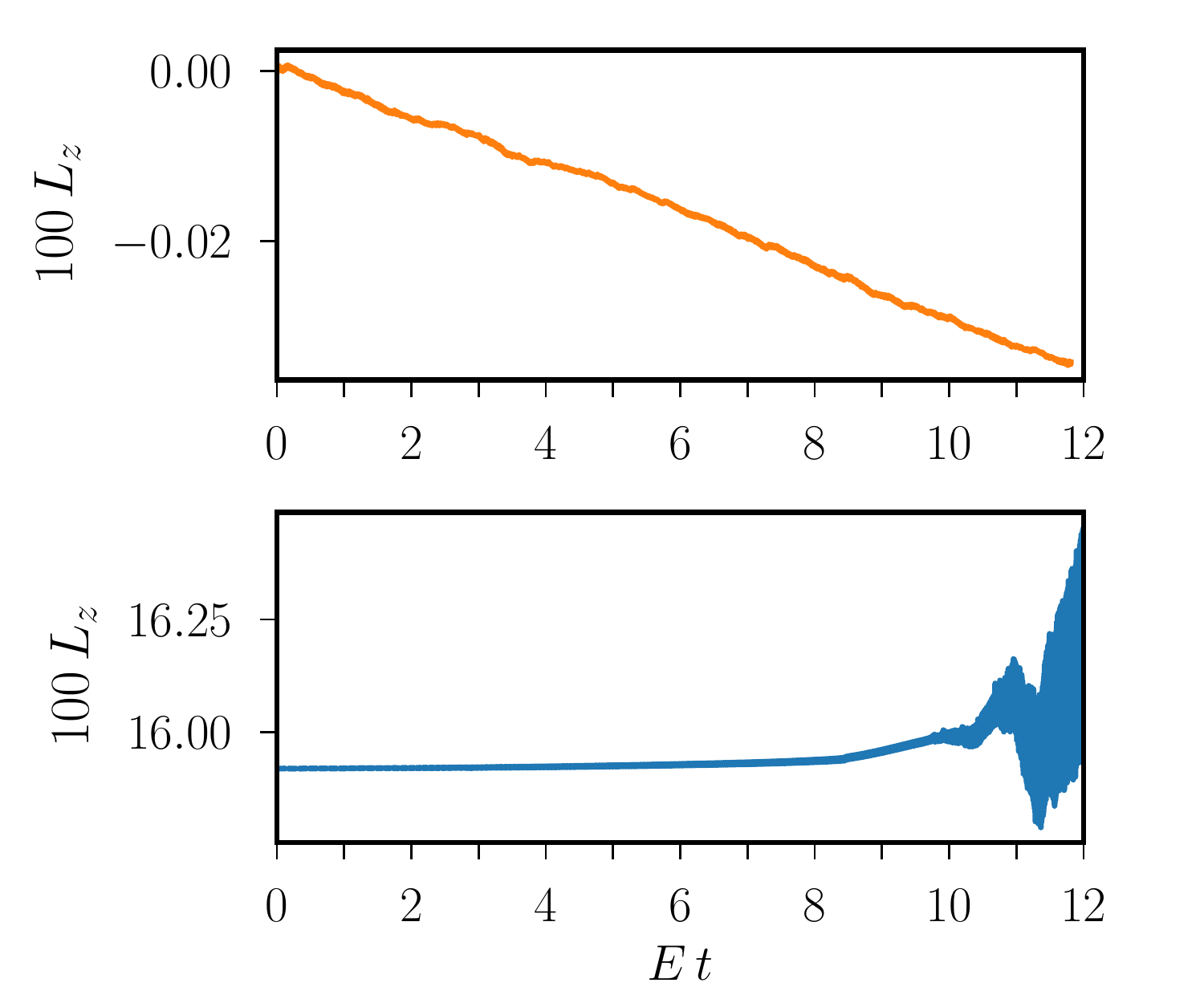} \\
    (a) Non-rotating  $\boldsymbol{\Omega} \simeq \boldsymbol{0}$ & (b) Rotating $\boldsymbol{\Omega} \propto \boldsymbol{1}_z$ \\
    \end{tabular}
    \caption{Non-convergence of the angular momentum $L_z$ in DNS after several viscous time units.
    Precession forcing given by definition (\ref{eq:deltaPrec}) with $P_x = 10^{-2}$ in stress-free spheroids ($a=b=1$, $c=0.95$). 
    At $t=0$, $[\omega_x,\omega_y]$ are chosen to match asymptotic solution (\ref{eq:PrecTriax1}). (a) DNS at $Po=-1$ for the two values of the Ekman number $E=5 \times 10^{-3}$  \citep[e.g. as considered in][]{wu2009dynamo} and $E=5 \times 10^{-4}$. At $t=0$, $\omega_z \approx 0$ for the two simulations. (b) DNS at $Po=-1.8$ and $E=5\times 10^{-4}$ for $\omega_z \approx 0$ (top panel) and $\omega_z = 0.1$ (bottom panel) at $t=0$.}
    \label{fig:derive}
\end{figure}

We can now inspect the long-term evolution of angular momentum since pathological behaviours have been reported in some axisymmetric configurations \citep{guermond2013remarks}. 
To illustrate this behaviour, we expand the angular momentum as $\boldsymbol{L}=\boldsymbol{L}_{0} + \boldsymbol{L}_{1}$, where $\boldsymbol{L}_{0}$ is the angular momentum of a dynamical solution of the problem and $\boldsymbol{L}_{1}$ is a modification of $\boldsymbol{L}_{0}$ associated with an additional uniform-vorticity flow. 
The time evolution of $\boldsymbol{L}_{1}$ is then given in the rotating frame by \citep[e.g.][]{roberts2012theory}
\begin{equation}
    \mathrm{d}_t \boldsymbol{L}_{1} + \boldsymbol{\Omega}_c \times \boldsymbol{L}_{1} = \boldsymbol{\Gamma}_{p,{1}} +  \boldsymbol{\Gamma}_{\nu,{1}}, 
    \label{eq:conservationL}
\end{equation} 
where $\boldsymbol{\Gamma}_{p,{1}} = \int_S p_{1} \, \boldsymbol{1}_n \times \boldsymbol{r} \ \mathrm{d} S$ is the pressure torque and $\boldsymbol{\Gamma}_{\nu,{1}}$ is the viscous torque. 
Since the viscous and pressure torques are non-zero when $a \neq b \neq c$, equation (\ref{eq:conservationL}) shows that the angular momentum is affected by viscosity in triaxial ellipsoids. 
The situation is possibly different in axisymmetric geometries. 
If the fluid is not globally rotating (i.e. when $\boldsymbol{\Omega} = \boldsymbol{0}$), then the component $\boldsymbol{L}_{1} \boldsymbol{\cdot} \boldsymbol{1}_i$ carried by the uniform-vorticity element $\boldsymbol{e}_i$ is arbitrary
when $\boldsymbol{1}_i$ is a revolution symmetry axis (since $\boldsymbol{\Gamma}_{p,{1}} \boldsymbol{\cdot} \boldsymbol{1}_i =  \boldsymbol{\Gamma}_{\nu,{1}} \boldsymbol{\cdot} \boldsymbol{1}_i = 0$). 
Similarly, if the fluid is globally rotating along the revolution symmetry axis $\boldsymbol{1}_i$, then the perturbation angular momentum $\boldsymbol{L}_{1} \propto \boldsymbol{1}_i$ is arbitrary \citep[it will depend on the initial conditions, e.g. as shown in][]{guermond2013remarks}.

The two situations are illustrated numerically in figure \ref{fig:derive} for a spheroid $a=b$ subject to the precession forcing (see its definition below in \S\ref{sec:precession}).
We have performed DNS using the standard finite-element method as implemented in the commercial software \textsc{comsol}.
The latter has already been employed to simulate precession-driven flows in ellipsoids with NS-BC \citep[e.g.][]{noir2013precession} and can also account for SF-BC \citep[e.g. for tidal flows in][]{cebron2013elliptical}.
The geometry is modelled by an unstructured mesh with tetrahedral elements in the bulk, surrounded by a boundary-layer mesh (made of prism elements) to ensure the convergence of the thin Ekman layer. 
We have employed Lagrange elements P$2$-P$3$ (i.e. quadratic for the pressure field and cubic for the velocity field). 
The total number of degrees of freedom ranges between $3 \times 10^5$ and $5 \times 10^5$, such that every targeted simulation took a few days to run in parallel on a cluster (to investigate the long-term evolution of $\boldsymbol{L}$). 
We observe that the axial angular momentum $L_z$ does not converge in time for the considered stress-free spheroid (it is still growing or decaying even after several viscous time scales) if either the fluid is non-rotating in average as in panel (a) or $\boldsymbol{\Omega} \propto \boldsymbol{1}_z$ as in panel (b). 
However, a definitive conclusion should not be drawn for every axisymmetric geometry. 
The situation is indeed different if the global rotation is not aligned with the revolution axis, since the three components of the angular momentum should be strongly coupled in equation (\ref{eq:conservationL}) for such configurations (even if $\boldsymbol{\Gamma}_\nu \boldsymbol{\cdot} \boldsymbol{1}_i = 0$, see \S\ref{sec:precession}).

\section{Application to precession-driven flows}
\label{sec:precession}
We consider precession-driven flows in ellipsoids, which have only received scant attention with SF-BC \citep{lorenzani2003inertial,wu2009dynamo,guermond2013remarks}. 
We work in the mantle frame rotating with respect to the inertial frame at the dimensionless angular velocity \citep[e.g.][]{noir2013precession}
\begin{subequations}
\label{eq:deltaPrec}
\begin{equation}
    \boldsymbol{\Omega}_c (t) = \underbrace{(1 + P_z)  \boldsymbol{1}_z}_{\boldsymbol{\Omega}} + \, \boldsymbol{\delta} (t), \quad \boldsymbol{\delta}(t) = P_x \left [ \cos(t) \boldsymbol{1}_x - \sin(t) \boldsymbol{1}_y  \right ],
    \tag{\theequation \emph{a,b}}
\end{equation}
\end{subequations}
with $P_x = Po \, \sin(\alpha)$ and $P_z = Po \, \cos(\alpha)$, where $Po = \Omega_p/\Omega_0$ is the Poincar\'e number ($\Omega_p$ being the angular velocity of precession and $\Omega_0$ that of the mantle) and $\alpha$ is the angle of precession measured from $\boldsymbol{1}_z$. 
Because the Poincar\'e force $\boldsymbol{r}\times \mathrm{d}_t \boldsymbol{\delta}$ is linear in Cartesian coordinates, the primary response of the fluid is a laminar uniform-vorticity flow \citep[e.g.][]{noir2013precession,kida2020prec}, on top of which secondary flows and turbulence can develop. 
For analytical progress, we expand the velocity field as $\boldsymbol{v}~=~\boldsymbol{v}_{0}+\boldsymbol{v}_{1}$, where $\boldsymbol{v}_{0}$ is the primary forced flow (which is mainly of uniform vorticity) and $\boldsymbol{v}_{1}$ represents small-amplitude additional flows such that $|\boldsymbol{v}_{1}| \ll |\boldsymbol{v}_{0}|$. 
We first seek analytical solutions of the primary flow in \S\ref{subsec:poincare}, which are compared with DNS in \S\ref{subsec:simus}.
Then, we explore the flow instabilities $\boldsymbol{v}_{1}$ growing upon the forced flow in \S\ref{subsec:instab}. 

\subsection{Laminar forced flows}
\label{subsec:poincare}
The forced laminar flows, which have been explored for a long time after the seminal work of \citet{poincare1910precession}, can be obtained using boundary-layer theory (BLT) in the low-viscosity regime $E \ll 1$ for SF-BC. 
To do so, we seek $\boldsymbol{v}_{0}$ as
\begin{equation}
    \boldsymbol{v}_{0}(\boldsymbol{r},t) \simeq \underbrace{\omega_x (t) \boldsymbol{e}_x + \omega_y (t) \boldsymbol{e}_y + \omega_z (t) \boldsymbol{e}_z}_{\boldsymbol{U}(\boldsymbol{r},t)} + E^{1/2} \widetilde{\boldsymbol{U}}(\boldsymbol{r},t)
    \label{eq:BLTGP1}
\end{equation}
where $\boldsymbol{U}(\boldsymbol{r},t)$ is a forced uniform-vorticity flow carrying angular momentum, and $\widetilde{\boldsymbol{U}}(\boldsymbol{r},t)$ is the viscous flow within the boundary layer at the leading order in $E^{1/2}$ \citep[e.g.][]{rieutord1992ekman}. 
A direct consequence of asymptotic expansion (\ref{eq:BLTGP1}) is that the bulk flow for SF-BC can be determined without explicitly solving for the boundary-layer flow (since the latter has an amplitude that is $E^{1/2}$ smaller than the bulk flow amplitude). 
The exact viscous torque given by formula (\ref{eq:viscoustorque1}) can then be approximated as
\begin{equation}
    \boldsymbol{\Gamma}_\nu \simeq -\frac{16 \pi}{3}abc \, E \, \mathrm{diag} \left [ \dfrac{(b^2-c^2)^2}{(b^2+c^2)^2}, \dfrac{(a^2-c^2)^2}{(a^2+c^2)^2}, \dfrac{(a^2-b^2)^2}{(a^2+b^2)^2} \right ] \, \boldsymbol{\omega}.
    \label{eq:viscoustorque2}
\end{equation}
The viscous flow $E^{1/2} \, \widetilde{\boldsymbol{U}}$ in expansion (\ref{eq:BLTGP1}) has a contribution of amplitude $\mathcal{O}(E^{3/2})$ to the viscous torque (since $|\boldsymbol{\epsilon} (\widetilde{\boldsymbol{U}})| = \mathcal{O}(E^{-1/2})$ and the volume scales as $\mathcal{O}(E^{1/2})$ within the Ekman layer), which can be neglected compared with expression (\ref{eq:viscoustorque2}) in the asymptotic regime $E \ll 1$. 
We recover from formula (\ref{eq:viscoustorque2}) that $\boldsymbol{\Gamma}_{\nu} \boldsymbol{\cdot} \boldsymbol{1}_i =0$ when the Cartesian vector $\boldsymbol{1}_i$ is a revolution symmetry axis, but also that the three components of the viscous torque are non-zero when $a \neq b \neq c$.
Then, the momentum equation reduces to
\begin{equation}
	\mathrm{d}_t \boldsymbol{\omega} - \left[  \left ( \boldsymbol{\omega} + \boldsymbol{\Omega}_c \right ) \boldsymbol{\cdot} \nabla \right ] \boldsymbol{U} = - \mathrm{d}_t \boldsymbol{\Omega}_c + \mathcal{L} \boldsymbol{\Gamma}_\nu,
	\label{eq:ODEGP1}
\end{equation}
where $\boldsymbol{\Gamma}_\nu$ is the viscous torque given by formula (\ref{eq:viscoustorque2}) and $\mathcal{L}$ is the matrix given by the inverse of expression (\ref{eq:Linv}b). 
The approximated viscous term is thus
\begin{equation}
  \mathcal{L}  \boldsymbol{\Gamma}_\nu = -5 E \mathrm{diag} \left [ \dfrac{(b/c-c/b)^2}{b^2+c^2}, \dfrac{(a/c-c/a)^2}{a^2+c^2}, \dfrac{(a/b-b/a)^2}{a^2+b^2} \right ] \, \boldsymbol{\omega}.
  \label{eq:LGamma}
\end{equation}
Equations (\ref{eq:ODEGP1}) and (\ref{eq:LGamma}) extend the asymptotic viscous model of \citet{noir2013precession} to stress-free ellipsoids, but we remind the reader that this stress-free model is not valid in spheres (since the angular momentum would be arbitrary in spheres because of $\boldsymbol{\Gamma}_\nu = \boldsymbol{0}$). 
The close similarity between the no-slip and stress-free cases, for which only the expression of the viscous term in equation (\ref{eq:ODEGP1}) differs, suggests that the same interior solution should be approached when $E \to 0$ in no-slip and stress-free ellipsoids.

Precession is often characterised by $|P_x| \ll 1$ in planetary liquid cores \citep[e.g.][]{noir2013precession}. 
Hence, we seek asymptotic solutions of equation (\ref{eq:ODEGP1}) in powers of $P_x$ as
\begin{equation}
    \boldsymbol{\omega} (t) = \boldsymbol{\omega}^{(0)} (t) + P_x \, \boldsymbol{\omega}^{(1)} (t) + P_x^2 \, \boldsymbol{\omega}^{(2)} (t) + \dots
\end{equation}
Since the mean rotation axis is $\boldsymbol{\Omega} \propto \boldsymbol{1}_z$ when $|P_x| \ll 1$, we assume that $a \neq b$ (to avoid the pathological situations outlined in \S\ref{sec:formulation} for the angular momentum conservation). 
The zeroth-order solution $\boldsymbol{\omega}^{(0)} (t)$ corresponds to a decaying transient when $t\to \infty$ (because of viscosity).
We thus discard $\boldsymbol{\omega}^{(0)} (t)$ in the following and solve the first-order problem in $P_x$. 
In the regime of vanishing viscosity $E \to 0$, we obtain the first-order solution 
\begin{subequations}
\allowdisplaybreaks
    \label{eq:PrecTriax1}
    \begin{equation}
        \omega_x^{(1)} (t) \simeq - \frac{1+[1+ P_z] A_1}{1-[1 + P_z]^2  \lambda_\text{so}^2} \cos(t), \quad 
        \omega_y^{(1)} (t) \simeq  \frac{1+[1+P_z] B_2}{1-[1 + P_z]^2 \lambda_\text{so}^2} \sin(t)
        \tag{\theequation \emph{a,b}}
    \end{equation}
\end{subequations}
and $\omega_z^{(1)} (t) \to 0$, with 
$A_1 = 2 a^2/(a^2+c^2)$, $B_2 = {2 b^2}/(b^2+c^2)$, and $\lambda_\text{so}=\sqrt{A_1 B_2}$. 
We have finally to compute the second-order solution $\boldsymbol{\omega}^{(2)}$, accounting for weakly nonlinear interactions in the viscous interior, to estimate the axial angular velocity (since it is undefined at the first order). 
An analytical solution can be obtained when $E \neq 0$, showing that $\boldsymbol{\omega}^{(2)} = \omega_z^{(2)} \boldsymbol{1}_z$, but the general expression of $\omega_z^{(2)}$ is too lengthy to be given here.
In the regime of vanishing viscosity $E \to 0$, it simplifies into
\begin{subequations}
\label{eq:meanflow}
\begin{equation}
    \omega_z^{(2)} (t) =  \frac{c^2}{4 \, \mathcal{D}_2^2}  \left [ \overline{\omega}_z^{(2)} + \delta \omega_z^{(2)} \cos(2t) \right ]
\end{equation}
with the denominator $\mathcal{D}_2 = a^2b^2 \widetilde{P}_z \left (P_z+1/2 \right ) -c^2 \left(a^2+b^2+c^2 \right)/4$ and $\widetilde{P}_z = P_z + 3/2$, where the amplitude of the mean geostrophic flow is given by
\begin{equation}
\overline{\omega}_z^{(2)} = - \left (\frac{c^2}{2} + a^2 \widetilde{P}_z \right ) \left (\frac{c^2}{2} + b^2 \widetilde{P}_z \right ) \frac{a^2+b^2}{(a/b-b/a)^2} \left [\left (\frac{a}{c}-\frac{c}{a} \right )^2 + \left (\frac{b}{c}-\frac{c}{b} \right )^2 \right ] 
\end{equation}
\end{subequations}
and that of the oscillatory component by $\delta \omega_z^{(2)} = (P_z+1)(a^2-b^2) ( a^2b^2 {\widetilde{P}_z}^2 - c^4/4 )$.
It is worth noting that the mean geostrophic flow $\overline{\omega}_z^{(2)}$ has an amplitude that is independent of $E$ in the vanishing regime $E \to 0$, which is somehow similar to the mean geostrophic flows driven by nonlinear boundary-layer interactions for NS-BC \citep[e.g.][]{cebron2021mean}.

A striking property of the asymptotic solution is that it exhibits two inviscid direct resonances, which occur when the common denominator in expressions (\ref{eq:PrecTriax1}a,b) vanishes at the two resonant values $Po^\pm$ given by $\lambda_\text{so} \left [ 1 + Po^\pm \cos(\alpha) \right ] = \pm 1$. 
The resonance associated with $Po^+$ actually corresponds to the inviscid resonance initially predicted by \citet{poincare1910precession}, which has been observed for no-slip boundaries
\citep[e.g.][]{vormann2018numerical,nobili2021hysteresis,burmann2022experimental}.
However, the second resonance at $Po^-$ is new, although precession-driven flows have been explored for more than a century in triaxial ellipsoids \citep[e.g.][]{poincare1910precession,noir2013precession}. 

\subsection{Numerical simulations}
\label{subsec:simus}
\begin{figure}
    \centering
    \begin{tabular}{cc}
    \begin{tabular}{c}
        \includegraphics[width=0.49\textwidth]{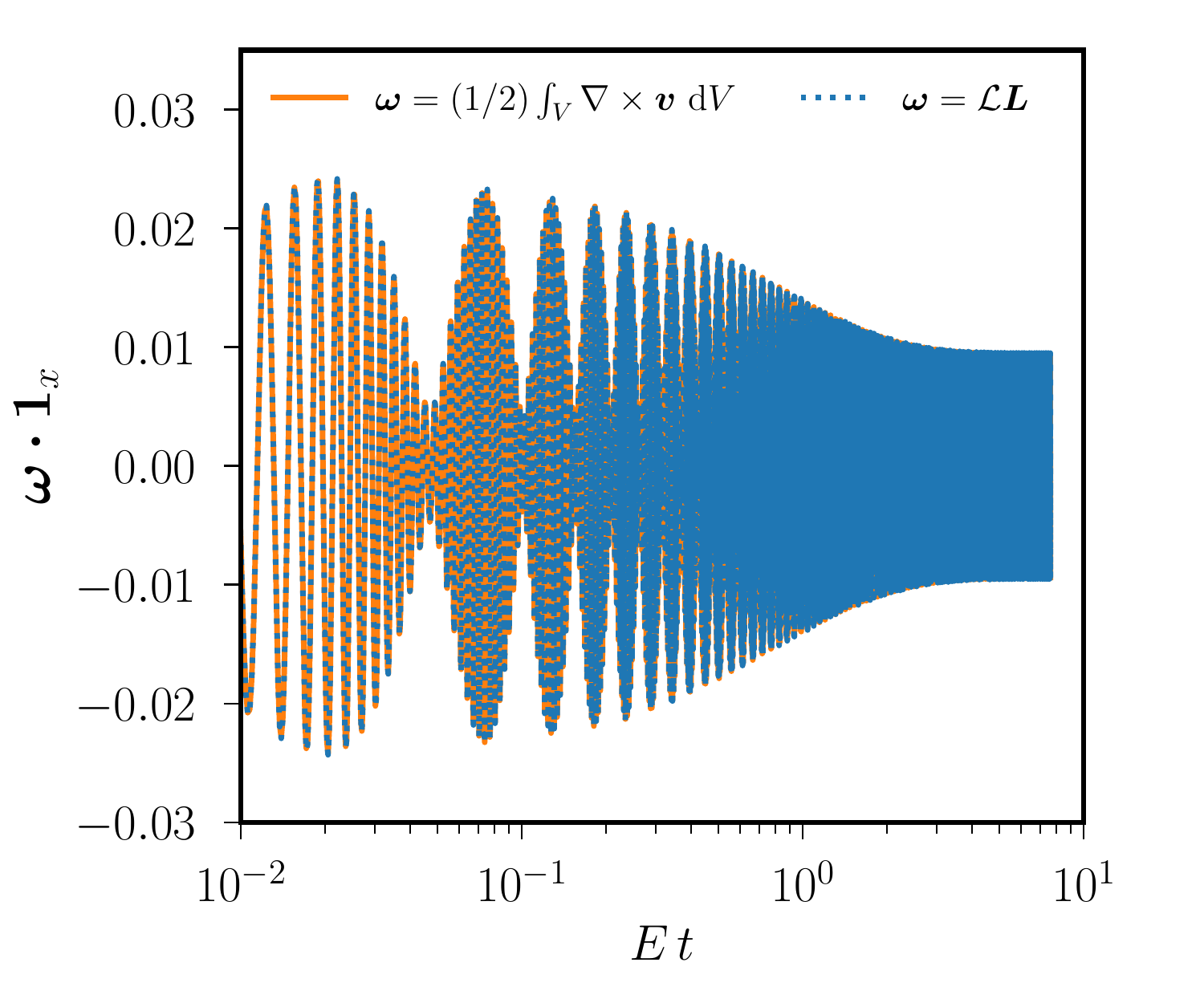}\\
    \end{tabular} 
    &
    \begin{tabular}{c}
        \includegraphics[width=0.49\textwidth]{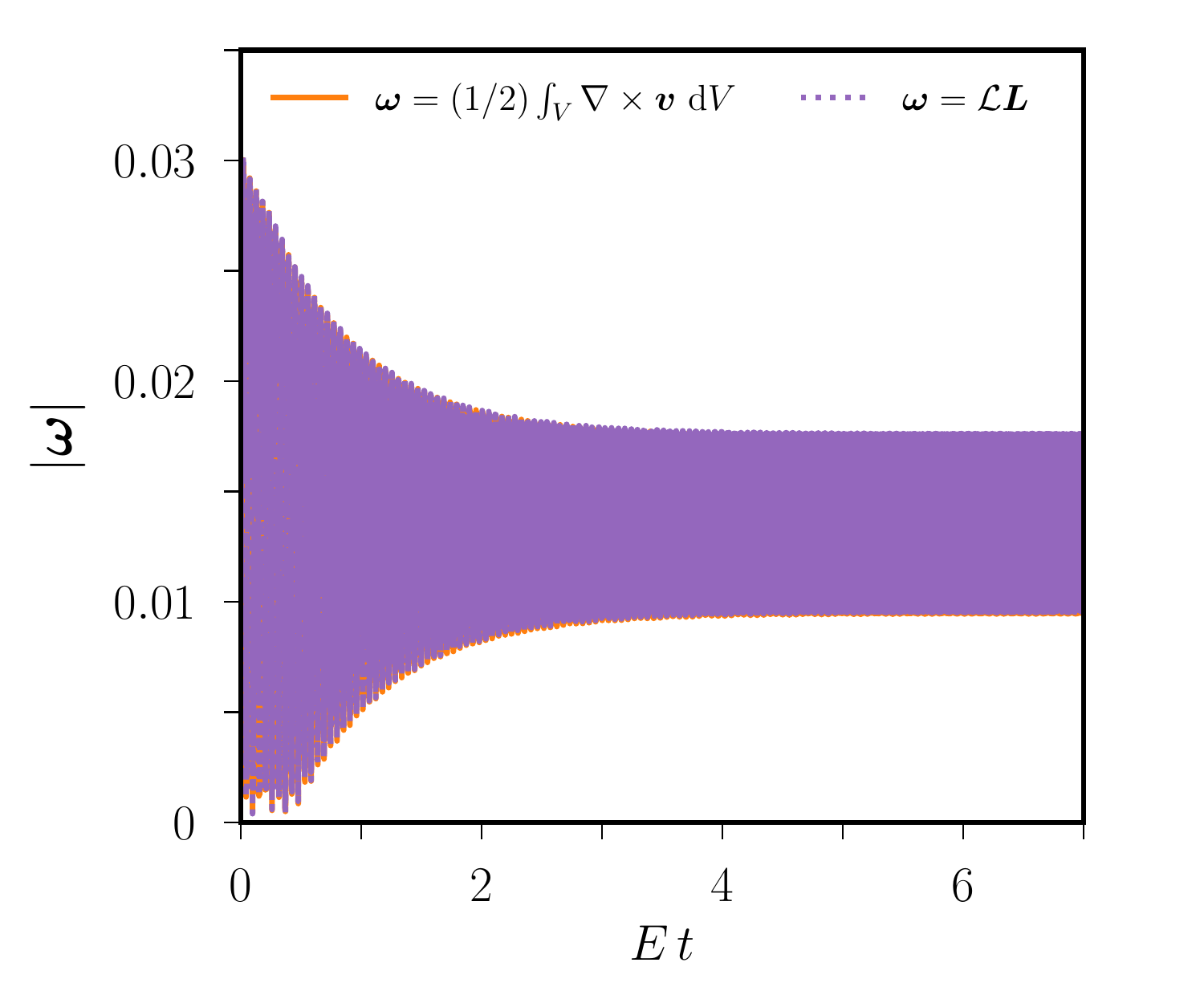}\\
    \end{tabular} \\
    (a) & (b) \\
    \end{tabular}
    \caption{DNS of precessing ellipsoids with SF-BC at $Po=-1.8$, $E=5 \times 10^{-4}$ and $P_x = Po \sin (\alpha) = 10^{-2}$. 
    Axisymmetric geometry $a=1.5$ and $b=c=1$.
    (a) Time evolution of the Cartesian component $\boldsymbol{\omega} \boldsymbol{\cdot} \boldsymbol{1}_x$ and (b) absolute value $|\boldsymbol{\omega}|$ of the angular velocity, computed in the DNS either from the volume-averaged vorticity as $\boldsymbol{\omega} = (1/2) \int_V \nabla \times \boldsymbol{v} \ \mathrm{d}V$ or using the angular momentum as $\boldsymbol{\omega} = \mathcal{L} \boldsymbol{L}$ using expression (\ref{eq:Linv}).}
    \label{fig:DNS}
\end{figure}

We have checked that the analytic expressions are in excellent agreement with the numerical integration of the exact uniform-vorticity model (\ref{eq:ODEGP1}) when $E \to 0$ (not shown).
Yet, it remains to confirm the validity of the asymptotic solutions against DNS with SF-BC. 
We first show in figure \ref{fig:DNS} the time evolution of the rotation vector $\boldsymbol{\omega} (t)$ in the DNS (performed with \textsc{comsol}, as explained in \S\ref{sec:formulation}). 
We illustrate the DNS at $P_x=10^{-2}$ with $Po=-1.8$ and $E=5 \times 10^{-4}$, in the particular axisymmetric geometry $a=1.5$ and $b=c=1$ (other parameters yield similar results, not shown). 
The fluid angular velocity $\boldsymbol{\omega}$ has been computed in the DNS using either the volume-averaged vorticity or formula (\ref{eq:Linv}a) after having computed the angular momentum. 
Both methods are found to be in excellent quantitative agreement for the SF-BC (as observed in the figure). 
For such an axisymmetric geometry, we may naively think (before any computation) that the long-term evolution of $\omega_x$ (or equivalently that of $L_x$) is unconstrained due to the vanishing component of the viscous torque $\boldsymbol{\Gamma}_\nu \boldsymbol{\cdot} \boldsymbol{1}_x = 0$ according to formula (\ref{eq:viscoustorque2}). 
We observe that $\omega_x$ initially displays a complicated transient (panel a), which dies out because of viscosity as expected from the asymptotic theory. 
Then, it converges towards a well-defined oscillatory state after a few viscous time scales (i.e. when $E \, t \gg 1$ in dimensionless units). 
The total angular velocity $\boldsymbol{\omega}$, which exhibits no long-term spurious dynamics (panel b), has a small amplitude compared with the mean rotation axis of the fluid $\boldsymbol{\Omega}=\boldsymbol{1}_z$ with respect to the inertial frame.  
We have checked that the final state is robust, as it is recovered by varying the numerical resolution and adopting different initial conditions for a few values of $Po$ and $E$ \citep [although multiple solutions may exist close to the inviscid resonances, as shown for sufficiently small Ekman numbers with NS-BC in][]{cebron2015bistable}. 

\begin{figure}
    \centering
    \begin{tabular}{cc}
    \includegraphics[width=0.49\textwidth]{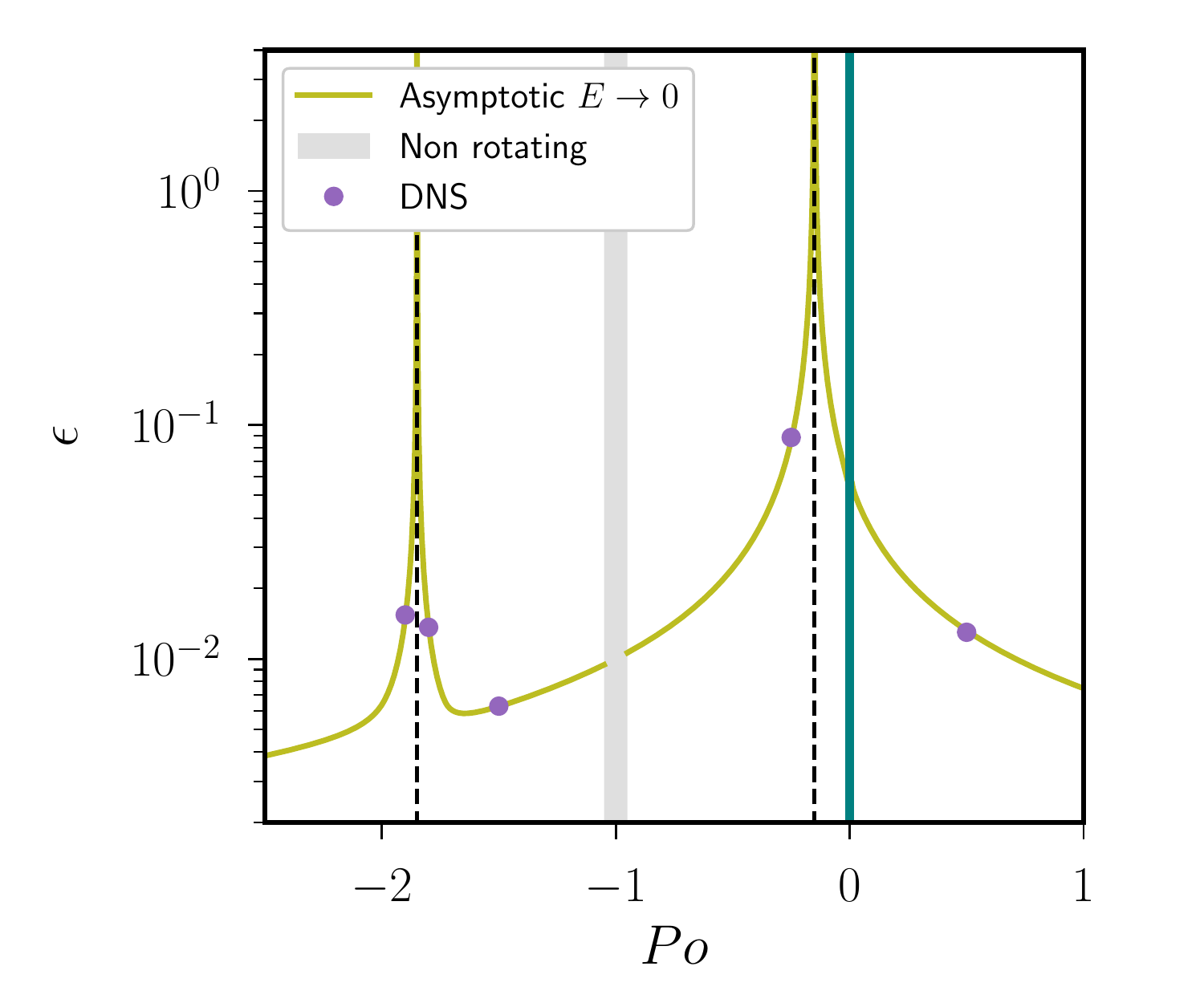} & 
    \includegraphics[width=0.49\textwidth]{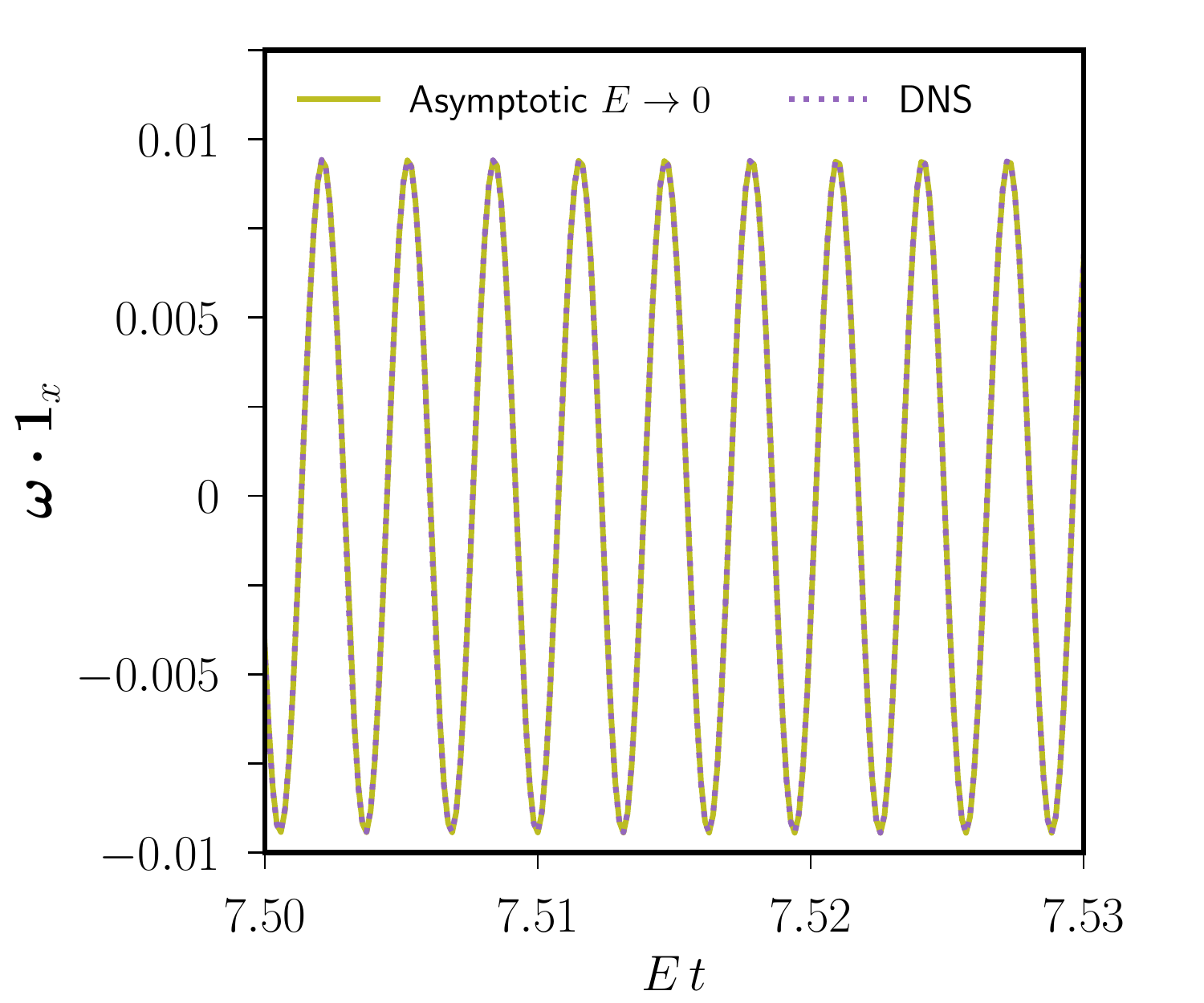} \\
    (a) & (b) \\
    \end{tabular}
    \caption{Precession-driven flows (SF-BC) at $P_x = Po \sin (\alpha) = 10^{-2}$ for $a=1.5$ and $b=c=1$. Comparison between asymptotic solution (\ref{eq:PrecTriax1}) and DNS at $E= 5 \times 10^{-4}$.
   (a) Time-averaged angular velocity $\epsilon = \overline{|\boldsymbol{\omega}|}$ as a function of $Po$. The fluid is not globally rotating when $Po \simeq -1$ if $|P_x| \ll 1$ (grey area). 
   Vertical dashed lines show the two resonances of asymptotic solutions (\ref{eq:PrecTriax1}) at $Po^\pm = - \sqrt{P_x^2 + (1/\lambda_\text{so}\mp1)^2}$. 
    Teal vertical line shows the region $|Po|<10^{-2}$ where no $\alpha$ can satisfy $P_x=10^{-2}$. 
    (b) Value of $|\boldsymbol{\omega}|$ as a function of the re-scaled time $E \, t$ at $Po=-1.8$.}
    \label{fig:benchmark}
\end{figure}

The comparison between the asymptotic results and the DNS is further illustrated in figure \ref{fig:benchmark}, still considering the illustrative axisymmetic geometry $a=1.5$ and $b=c=1$ (other geometries with $a \neq b$ give again similar results, not shown). 
The DNS are in excellent quantitative agreement with the asymptotic solution, although the latter has been obtained assuming $E \to 0$, for both the time-averaged and the instantaneous angular velocity (see panel b after seven viscous time scales). 
We also have checked that $\delta \omega_z^{(2)}$ is accurately recovered in the DNS (not shown). 
The observed excellent quantitative agreement with theoretical precession-driven flows has not been obtained using NS-BC in ellipsoids, both in DNS \citep[e.g.][]{noir2013precession} and laboratory experiments \citep[e.g.][]{nobili2021hysteresis,burmann2022experimental}. 
Finally, the DNS also confirm the physical existence of the two inviscid resonances of solutions (\ref{eq:PrecTriax1}).

\subsection{Asymptotic theory of flow instabilities}
\label{subsec:instab}
The forced laminar flow $\boldsymbol{U} (\boldsymbol{r},t)$, given by equation (\ref{eq:ODEGP1}) when $E \ll 1$, can be destabilised by various hydrodynamic instabilities in ellipsoids.
Precession-driven instabilities are classified either as viscously driven if they only exist when $E \neq 0$, or as inertial if they survive when $E = 0$. 
Viscous instabilities exist in no-slip spheres, such as boundary-layer instabilities \citep[e.g.][]{lorenzani2001fluid,buffett2021conditions} or the conical-shear instability \citep[e.g.][]{lin2015shear,cebron2019precessing}.
On the contrary, the inertial instabilities only exist in non-spherical geometries \citep[e.g.][]{kerswell1993instability,wu2011high,vidal2017odei}.
In the following, we extend the prior inviscid linear analyses of the inertial instabilities, which all considered precession at $\alpha=\pi/2$ and in the precession frame (i.e. only for spheroids), to account for the SF-BC and the time-dependent background flow (\ref{eq:PrecTriax1}) in the mantle frame. 
To do so, we expand the governing equations with respect to $\boldsymbol{U}$ (discarding the small-amplitude viscous flow $E^{1/2} \widetilde{\boldsymbol{U}}$ in the bulk, which is negligible when $E \ll 1$ as found in the DNS). 
The perturbation velocity $\boldsymbol{v}_{1}$, which is assumed to be of small amplitude compared with $\boldsymbol{U}$, is governed in the mantle frame by
\begin{subequations}
\label{eq:linearisedNS}
\begin{align}
 	\partial_t \boldsymbol{v}_{1} + 2 \boldsymbol{\Omega}_c \times \boldsymbol{v}_{1} &= \boldsymbol{\mathcal{L}} (\boldsymbol{v}_{1}) + 2 E \nabla \boldsymbol{\cdot} \boldsymbol{\epsilon}(\boldsymbol{v}_{1}) - \nabla p, \\
	\nabla \boldsymbol{\cdot} \boldsymbol{v}_{1} &= 0,
\end{align}
\end{subequations}
with the linearised advection operator $\boldsymbol{\mathcal{L}} (\boldsymbol{a}) = -( \boldsymbol{a} \boldsymbol{\cdot} \nabla ) \, \boldsymbol{U} - ( \boldsymbol{U} \boldsymbol{\cdot} \nabla ) \, \boldsymbol{a}$. 
The perturbation velocity $\boldsymbol{v}_{1}$ then satisfies the SF-BC \citep{mason2002chaotic,wu2009dynamo}
\begin{subequations}
\label{eq:BCSF2}
\begin{equation}
    \left . \boldsymbol{v}_{1} \boldsymbol{\cdot} \boldsymbol{1}_n \right |_S = 0, \quad \left . [ \boldsymbol{\epsilon} (\boldsymbol{v}_{1})  \boldsymbol{\cdot} \boldsymbol{1}_n] \times \boldsymbol{1}_n \right |_S = \boldsymbol{0}.
    \tag{\theequation \emph{a,b}}
\end{equation}
\end{subequations}
To explore the low-viscosity regime $E \ll 1$, which is difficult to probe using DNS, we develop an asymptotic model. 
We seek $\boldsymbol{v}_{1}$ using BLT as \citep[e.g.][]{rieutord1992ekman}
\begin{subequations}
\label{eq:expansionPrecBL}
\begin{equation}
    \boldsymbol{v}_{1} (\boldsymbol{r},t) \simeq \boldsymbol{u} (\boldsymbol{r},t) + E^{1/2} \, \widetilde{\boldsymbol{u}} (\boldsymbol{r},t), \quad \nabla \boldsymbol{\cdot} \boldsymbol{u} = 0, \quad \left . \boldsymbol{u} \boldsymbol{\cdot} \boldsymbol{1}_n \right |_S = 0,
    \tag{\theequation \emph{a--c}}
\end{equation}
\end{subequations} 
where $\boldsymbol{u}(\boldsymbol{r},t)$ represents the inviscid bulk flow and $\widetilde{\boldsymbol{u}}(\boldsymbol{r},t)$ is the leading-order viscous flow within the Ekman layer to satisfy SF-BC (\ref{eq:BCSF2}).
Because the boundary-layer flow has an amplitude that is $E^{1/2}$ smaller than the bulk flow amplitude, SF-BC strongly weaken the viscous instabilities in ellipsoids. 
In particular, the critical shear layers spawned by the Ekman layer at the critical latitudes are almost suppressed in stress-free ellipsoids without an inner core \citep{tilgner1999non}.
Consequently, the inertial instabilities triggered in the (nearly) inviscid bulk are expected to be largely favoured in stress-free ellipsoids (compared with viscous instabilities).  

To solve problem (\ref{eq:expansionPrecBL}), we introduce the finite-dimensional polynomial vector space $\boldsymbol{\mathcal{V}}_n$ spawned by the global real-valued incompressible elements $\{ \boldsymbol{u}_k\}$, made of Cartesian monomials $x^i y^j z^k$ of maximum degree $i+j+k \leq n$ and satisfying the no-penetration BC \citep[e.g.][]{vidal2020compressible,vidal2021acoustic}.
Such vector elements are indeed known to form a complete basis for smooth velocity fields in ellipsoids when $n \to \infty$ \citep[e.g.][]{lebovitz1989stability,backus2017completeness}. 
Then, we seek the bulk flow using the Galerkin expansion (written using Einstein's convention)
\begin{subequations}
\label{eq:GPexpansion}
\begin{equation}
    \boldsymbol{u}(\boldsymbol{r},t) = \alpha_k (t) \boldsymbol{u}_k (\boldsymbol{r}), \quad \nabla \boldsymbol{\cdot} \boldsymbol{u}_k = 0, \quad \left . \boldsymbol{u}_k \boldsymbol{\cdot} \boldsymbol{1}_n \right |_S = 0,
    \tag{\theequation \emph{a--c}}
\end{equation}
\end{subequations}
where $\boldsymbol{\alpha} = (\alpha_1, \alpha_2, \dots, \alpha_N)^\top$ is the state vector of the modal coefficients.
The number of elements $N$ for a given maximum degree $n$ in expansion (\ref{eq:GPexpansion}) is $N=n(n+1)(2n+7)/6$. 
In practice, we truncate the polynomial expansion at the maximum degree $n$, substitute the truncated expansion into equation (\ref{eq:linearisedNS}) and, finally, project the resulting equations onto every basis element $\boldsymbol{u}_i$ to minimise the residual with respect to the real-valued inner product defined by $\langle \boldsymbol{a}, \boldsymbol{b} \rangle_V = \int_V \boldsymbol{a} \boldsymbol{\cdot} \boldsymbol{b} \ \mathrm{d} V$. 
The governing equations then reduce to
\begin{equation}
    \boldsymbol{M} \,\mathrm{d}_t \boldsymbol{\alpha} = \left ( \boldsymbol{L} - \boldsymbol{C} - \boldsymbol{D} \right ) \boldsymbol{\alpha},
    \label{eq:EQSstabu}
\end{equation}
where $\boldsymbol{M}_{ij} = \langle \boldsymbol{u}_i, \boldsymbol{u}_j \rangle_V$ is the mass matrix, $\boldsymbol{C}_{ij} = \langle \boldsymbol{u}_i, 2 \boldsymbol{\Omega}_c \times \boldsymbol{u}_j \rangle_V$ represents the Coriolis force, $\boldsymbol{L}_{ij} = \langle \boldsymbol{u}_i, \boldsymbol{\mathcal{L}} (\boldsymbol{u}_j) \rangle_V$ is the matrix representing the linearised advection terms and the viscous matrix $\boldsymbol{D}$ is given by (after integration by parts)
\begin{equation}
    \boldsymbol{D}_{ij} = 2 E \int_V \boldsymbol{\epsilon} (\boldsymbol{u}_i) : \boldsymbol{\epsilon} (\boldsymbol{u}_j) \, \mathrm{d} V
    \label{eq:viscousproj}
\end{equation}
in which we have enforced SF-BC (\ref{eq:BCSF2}) in the projection to simplify the integration \citep[e.g.][]{guermond2013remarks}.
As already noticed for the forced flow, a useful consequence of expansion (\ref{eq:expansionPrecBL}) is that the bulk flow $\boldsymbol{u}$ can be determined in equation (\ref{eq:EQSstabu}) without an explicit solution of $\widetilde{\boldsymbol{u}}$ for SF-BC.
This has also been reported for asymptotic models of thermal convection or waves in rotating stress-free spheres \citep{liao2001viscous,zhang2004new}. 
This is a noticeable difference from asymptotic models using NS-BC, which require a matching between the boundary-layer flow and the interior solution \citep[which are of the same order of magnitude, e.g.][]{zhang2007asymptotic,zhang2014precessing}.

\begin{figure}
    \centering
    \begin{tabular}{c}
        \includegraphics[width=0.98\textwidth]{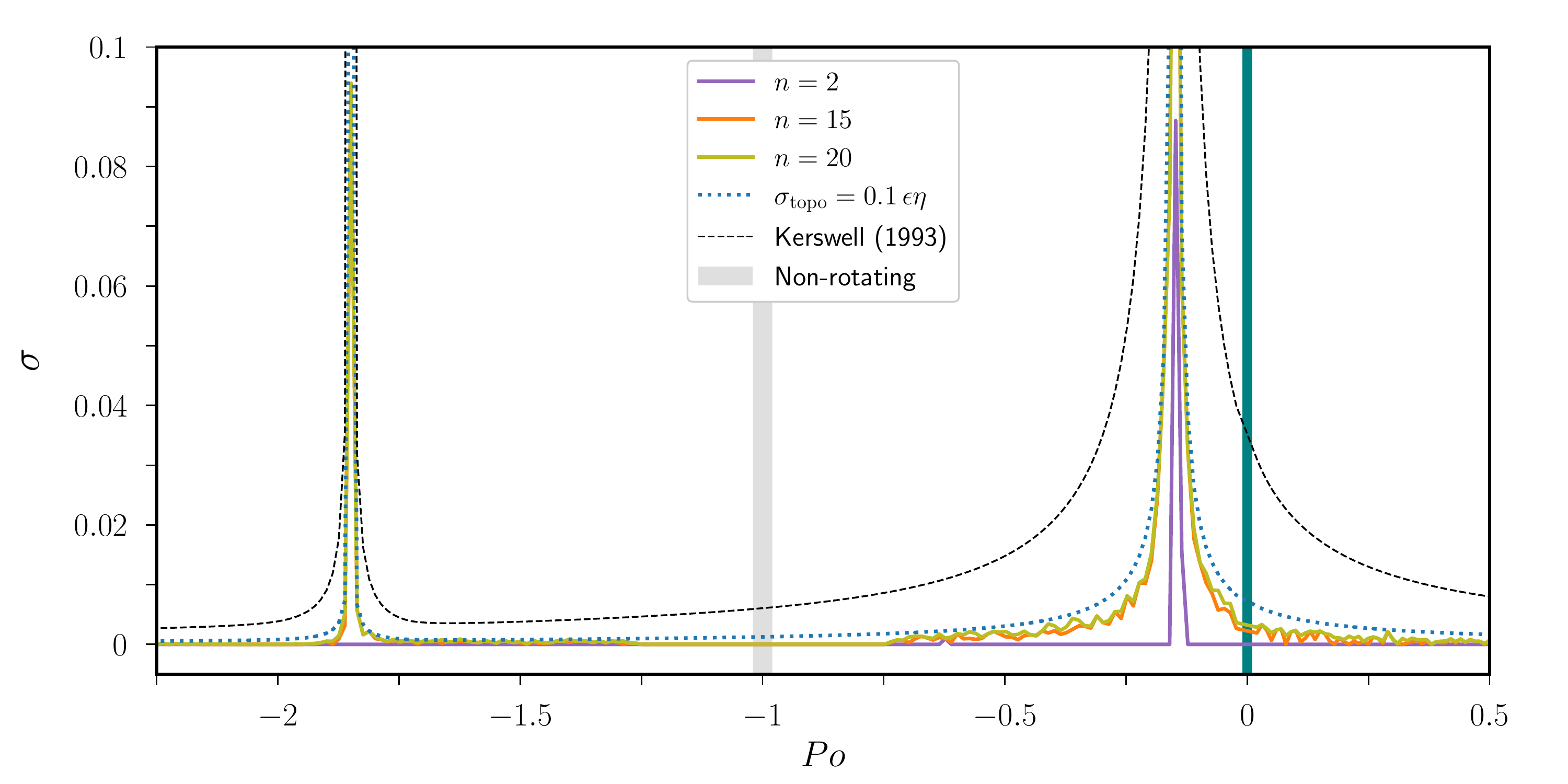} \\
        \includegraphics[width=0.98\textwidth]{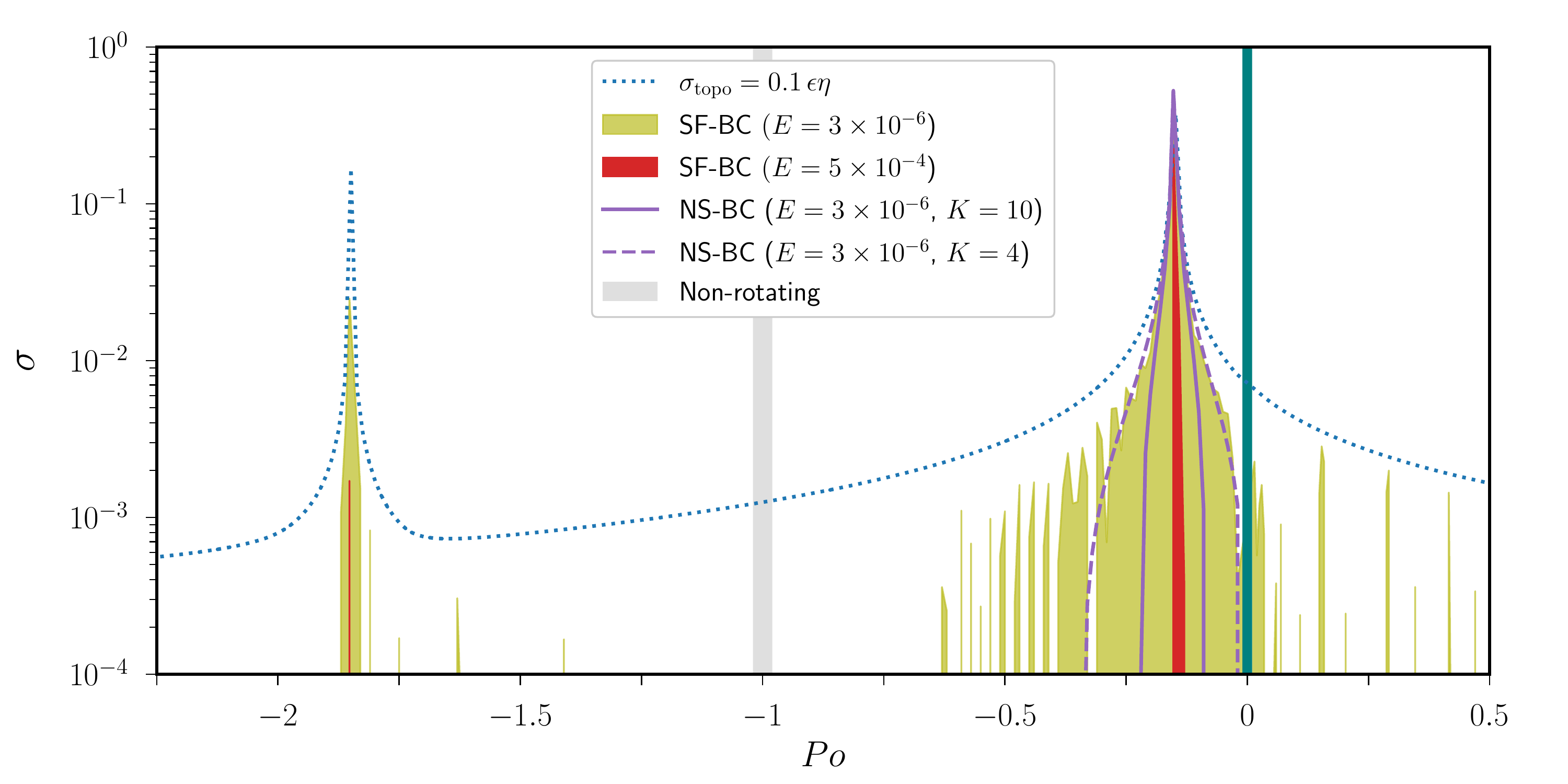} \\
    \end{tabular}
    \caption{Growth rate $\sigma$ of the inertial (topographic) instabilities growing upon flow (\ref{eq:PrecTriax1}) at $P_x=Po \sin(\alpha)=10^{-2}$, as a function of $Po$ (using sampled values). Teal vertical line shows the interval $|Po|<10^{-2}$ in which no $\alpha$ can satisfy $P_x=10^{-2}$. The fluid is not globally rotating near $Po \simeq -1$ when $|P_x| \ll 1$ (grey area). 
    (a) Inviscid growth rate for various degrees $n$ of the global modes. Dashed black curve is obtained in the unbounded short-wavelength limit \citep{kerswell1993instability}. 
    (b) Viscous effects. Dotted blue line shows the upper bound of the inviscid growth rate. Olive coloured area shows the unstable region for SF-BC at $E=3 \times 10^{-6}$, and thick red line shows the unstable zone for SF-BC at $E=5 \times 10^{-4}$ (both computed at $n=20$). Purple coloured curves show viscous growth rate (\ref{eq:sigmaNS}) for NS-BC, with $K \in [4,10]$ to account for the Ekman damping of the large-scale modes (see figure \ref{fig:ReGImG}).}
    \label{fig:floquet}
\end{figure}

Since asymptotic solution (\ref{eq:PrecTriax1}) is periodic of period $T=2\pi$, we investigate the linear stability using Floquet theory. 
We first compute the eigenvalues $\chi$ of the monodromy matrix $\boldsymbol{\Phi}(2\pi)$ given by
\begin{subequations}
\label{eq:floquet}
\begin{equation}
    \boldsymbol{M} \, \mathrm{d}_t \boldsymbol{\Phi} = ( \boldsymbol{L} - \boldsymbol{C} - \boldsymbol{D} ) \, \boldsymbol{\Phi}
    , \quad \boldsymbol{\Phi}(0) = \boldsymbol{I},
    \tag{\theequation \emph{a,b}}
\end{equation}
\end{subequations}
where $\boldsymbol{I}$ is the identity matrix.
Then, we compute the complex-valued Lyapunov exponents as $\mu = (1/T) \log \chi$ whose real part $\mathrm{Re}(\mu) = \sigma$ is the growth rate of the instability.
As initially noticed by \citet{kerswell1993instability} and \citet{wu2011high}, the finite-dimensional polynomial is left invariant by the linear operator in the momentum equation, that is $\boldsymbol{\mathcal{L}}(\boldsymbol{\mathcal{V}}_n) \in \boldsymbol{\mathcal{V}}_n$. 
Therefore, we can construct exact polynomial solutions of equation (\ref{eq:linearisedNS}) giving sufficient conditions for linear instability in the inviscid regime $E=0$. 

We show in figure \ref{fig:floquet}(a) the results of the linear inviscid stability analysis at $P_x = 10^{-2}$.
We have numerically solved equation (\ref{eq:floquet}) using a fourth-order Runge-Kutta solver and standard linear algebra routines. 
As in \citet{kerswell1993instability} and \citet{wu2011high}, there are no instabilities associated with the linear elements $n=1$. 
The first instabilities, which are here associated with the quadratic modes with $n=2$, only occur near the resonance at $Po^+$. 
When $n$ is increased, additional tongues of inertial (topographic) instabilities appear with a growth rate scaling in the inviscid regime as
\begin{equation}
    \sigma_\mathrm{topo} = \mathcal{O}(\epsilon \eta),
    \label{eq:sigmaTopo}
\end{equation}
where $\epsilon = \overline{|\boldsymbol{\omega}|}$ is the mean value of the differential rotation between the fluid and the mantle and $\eta = a^2/c^2 - 1$ is the polar flattening.
The numerical prefactor is found to be $ \sigma_\text{topo} /(\epsilon \eta) \approx 0.1$ when $n \leq 20$ (as shown in the figure). 
Moreover, when $n \to \infty$, the growth is expected to approach the upper bound given in the unbounded short-wavelength approximation \citep{kerswell1993instability}. 
This shows that the forced laminar flow is generically unstable to short-wavelength perturbations without viscosity. 

However, the short-wavelength modes are more damped by viscosity than the large-scale ones.
Consequently, viscous effects will select the allowable unstable modes for a given value of the Ekman number. 
To show this, we have explored the linear stability including viscous damping in figure \ref{fig:floquet}(b). 
At $E=5 \times 10^{-4}$, the forced flow is only unstable in extremely thin tongues near the two resonances at $Po^\pm$ for the SF-BC. 
This is consistent with the absence of instabilities in the DNS performed at $E=5 \times 10^{-4}$ (see figure \ref{fig:benchmark}). 
More challenging DNS with SF-BC at smaller values $E = \mathcal{O}(10^{-6})$, which are beyond the scope of the present paper, could allow us to obtain instabilities for values of $Po$ in a larger interval.
Finally, it is also useful to compare the stability of the forced flow with SF-BC and NS-BC. 
A proper asymptotic theory for the no-slip case, rooted in the BLT of the inertial modes \citep[e.g.][]{greenspan1968theory}, will be considered elsewhere.
Nonetheless, an upper bound for the viscous growth rate of the inertial instabilities can be estimated as
\begin{equation}
    \sigma_\text{topo} \approx 0.1 \, \epsilon \eta - K \sqrt{E [1 + P_z]},
    \label{eq:sigmaNS}
\end{equation}
assuming that the fluid is rotating on average at $1+P_z$ in the mantle frame. 
Here, the numerical prefactor $K = 4 - 10$ heuristically accounts for the Ekman damping of the large-scale flow structures with NS-BC (see figure \ref{fig:ReGImG}).
For the small value $E = 3 \times 10^{-6}$, we observe that the forced flow at $P_x = 10^{-2}$ would be mainly stable with NS-BC (except near the resonance $Po^+$), whereas it would be unstable for other values of $Po$ with SF-BC. 
Therefore, the figure clearly illustrates that adopting SF-BC (instead of NS-BC) can be useful to explore the turbulence driven by inertial instabilities in the bulk of the fluid.

\section{Discussion}
\label{sec:discussion}
\subsection{Physical insight from the Coriolis eigenmodes}
We have illustrated with the case of precession-driven flows that the long-term evolution of angular momentum is damped by viscosity in triaxial ellipsoids.
Similarly, viscosity affects the angular momentum in axisymmetric rotating ellipsoids if the mean rotation axis $\boldsymbol{\Omega}$ is not aligned with the revolution symmetry axis (even if $\boldsymbol{\Gamma}_i \boldsymbol{\cdot} \boldsymbol{1}_i = 0$ in such geometries, where $\boldsymbol{1}_i$ is the revolution axis along one of the principal semi-axes). 
Asymptotic analysis offers a physical understanding of why the cases $\boldsymbol{\Omega} \propto \boldsymbol{1}_i$ and $\boldsymbol{\Omega} \not \propto \boldsymbol{1}_i$ strongly differ in axisymmetric ellipsoids. 

When $E \ll 1$, the solutions of equations (\ref{eq:NSincomp}a,b) in stress-free or no-slip ellipsoids can be rigorously expanded onto a combination of the inviscid eigenmodes of the (steady) Coriolis operator given by \citep[e.g.][]{backus2017completeness} 
\begin{subequations}
\label{eq:IMpb}
\begin{equation}
    \mathrm{i} \lambda_k \nabla \times  \boldsymbol{Q}_k = -2 \nabla \times ( \boldsymbol{\Omega} \times \boldsymbol{Q}_k), \quad  \nabla \boldsymbol{\cdot} \boldsymbol{Q}_k = 0, \quad \left . \boldsymbol{Q}_k \boldsymbol{\cdot} \boldsymbol{1}_n \right |_S = 0,
    \tag{\theequation \emph{a--c}}
\end{equation}
\end{subequations}
where $[\lambda_k, \boldsymbol{Q}_k (\boldsymbol{r})]$ is the $k$th eigenvalue-eigenfunction pair. 
Only three of these eigenmodes carry a non-zero angular momentum in ellipsoids \citep[by virtue of the orthogonality of the eigenmodes, see][]{ivers2017enumeration}, namely the spin over mode $\boldsymbol{Q}_\text{so}$, its complex conjugate $\boldsymbol{Q}_\text{so}^\dagger$ and the zero-frequency geostrophic mode $\boldsymbol{Q}_\text{sup}$ associated with axial (differential) rotation along $\boldsymbol{\Omega}$.
Because these three modes are uniform-vorticity flows such as $\boldsymbol{Q}_k = \omega_{k,x} \, \boldsymbol{e}_x + \omega_{k,y} \, \boldsymbol{e}_y + \omega_{k,z} \, \boldsymbol{e}_z$, they are given by the matrix eigenvalue problem
\begin{equation}
    \begin{pmatrix}
    0 & {2 a^2 \Omega_z}/{(a^2+c^2)}& -{2 a^2 \Omega_y}/{(a^2+b^2)} \\ 
     -{2 b^2 \Omega_z}/{(b^2+c^2)} & 0 & {2 b^2 \Omega_x}/{(a^2+b^2)} \\
     {2 c^2 \Omega_y}/{(b^2+c^2)} & -{2 c^2 \Omega_x}/{(a^2+c^2)} & 0 \\
    \end{pmatrix} \boldsymbol{\omega}_k = \mathrm{i} \lambda_k \boldsymbol{\omega}_k
    \label{eq:matrixGP1}
\end{equation}
with $\boldsymbol{\Omega} = (\Omega_x, \Omega_y, \Omega_z)^\top$, where the rotation vector $\boldsymbol{\omega}_k = (\omega_{k,x},\omega_{k,y},\omega_{k,z})^\top$ of the eigenmode $\boldsymbol{Q}_k$  is given by the $k$th eigenvector of matrix (\ref{eq:matrixGP1}).  
Consequently, the uniform-vorticity components $\omega_i (t) \, \boldsymbol{e}_i$ of the flow in expansion (\ref{eq:expansionBuffett}) are not mutually independent in rotating ellipsoids but, instead, are tied to the dynamics of these modes.
More precisely, the equatorial components of the angular momentum $\boldsymbol{L} \times \boldsymbol{1}_\Omega$ are coupled through the dynamics of the two spin-over modes.  Similarly, the axial angular momentum $\boldsymbol{L} \boldsymbol{\cdot} \boldsymbol{1}_\Omega$ (related to the fluid spin-up) is piloted by the dynamics of the geostrophic mode $\boldsymbol{Q}_\text{sup}$. 

\begin{figure}
    \centering
    \begin{tabular}{cc}
        \includegraphics[width=0.49\textwidth]{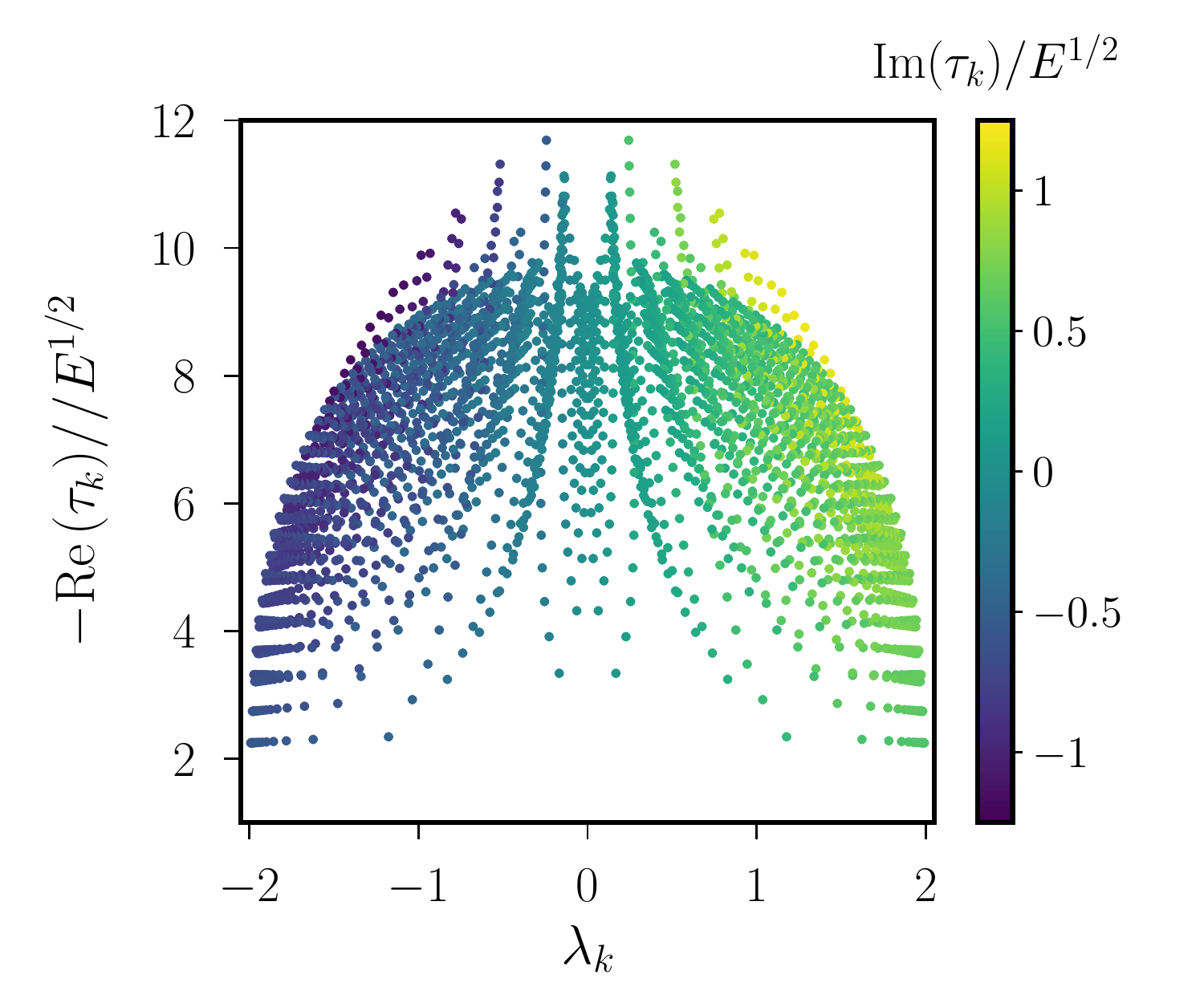} &
        \includegraphics[width=0.49\textwidth]{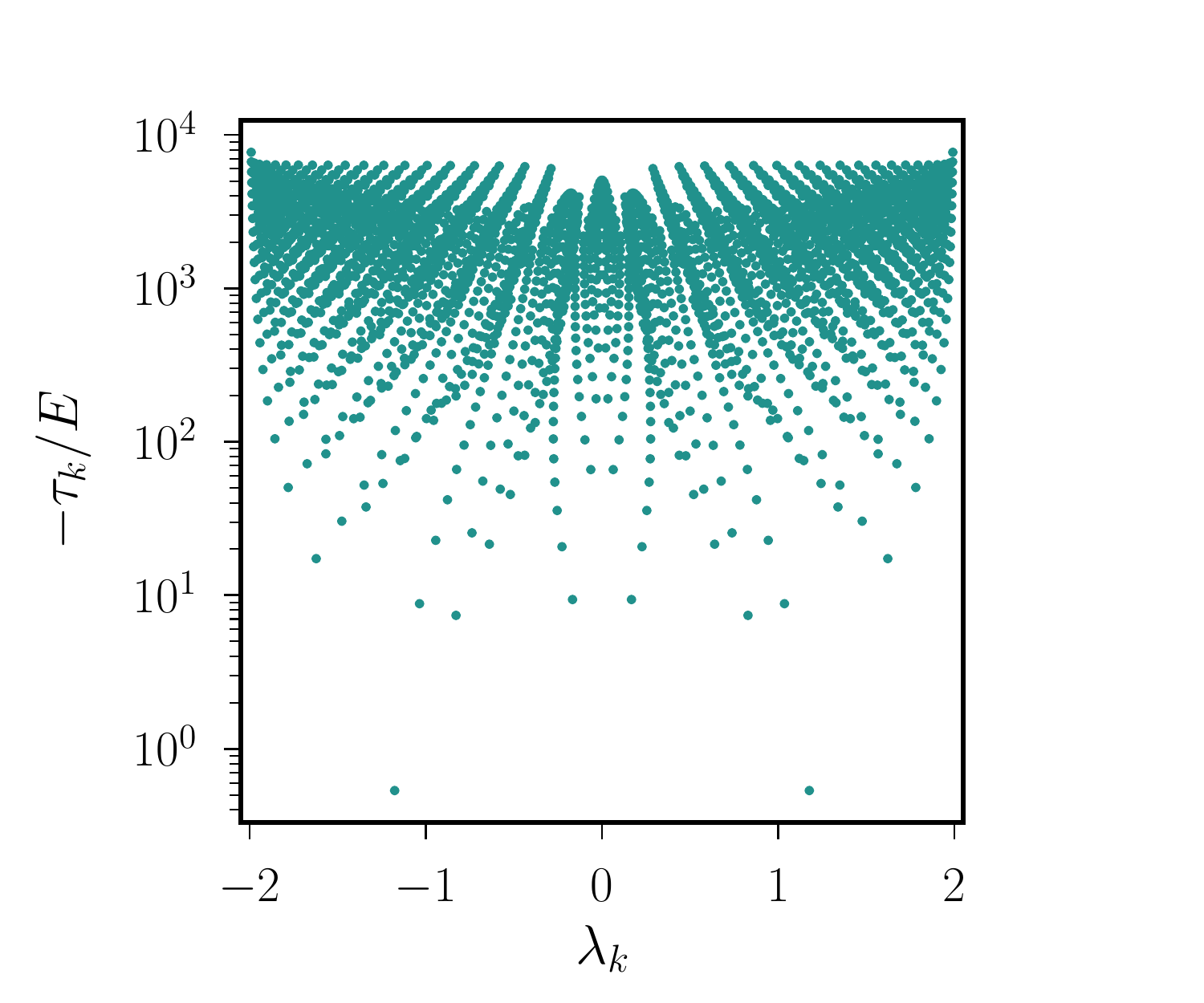} \\
        (a) NS-BC & (b) SF-BC \\
    \end{tabular}
    \caption{Viscous decay rates $\tau_k$ of the inertial modes of maximum polynomial degree $n=20$, as a function of the inviscid eigenfrequency $\lambda_k \neq 0$. Axisymmetric ellipsoid with semi-axes $a=1.5$ and $b=c=1$ rotating at the angular frequency $\boldsymbol{\Omega}=\boldsymbol{1}_z$. (a) Complex-valued $\tau_k$ for NS-BC. Colour bar shows the normalised imaginary part $\mathrm{Im}(\tau_k)$. (b) Real-valued $\tau_k$ (i.e. $\mathrm{Im}(\tau_k)=0$) given by formula (\ref{eq:decaySFVol}) for SF-BC.}
    \label{fig:ReGImG}
\end{figure}

From a physical viewpoint, whether viscosity affects the long-term evolution of angular momentum or not is thus deeply rooted in the viscous dynamics of these three eigenmodes. 
We can quantify how viscosity impacts the inviscid eigenmodes by estimating the global viscous decay rates $\tau_k$ of the Coriolis modes as
\begin{equation}
        \left . \partial_{t} \boldsymbol{Q}_k \right |_{t=0} \simeq \tau_k \, \boldsymbol{Q}_k.
\end{equation}
For NS-BC, $\tau_k$ is a complex-valued quantity with a real part $\mathrm{Re}(\tau_k) \leq 0$ representing the volume-averaged viscous decay rate, and an imaginary part $\mathrm{Im}(\tau_k)$ characterising the frequency shift due to viscous effects \citep[e.g.][]{greenspan1968theory}.
Typical values are illustrated in figure \ref{fig:ReGImG} for a particular ellipsoidal geometry. 
It has also been recognised for a long time that, for NS-BC (\ref{eq:BCNS}), the viscous torque in the mantle frame is related to the viscous damping of these three eigenmodes \citep[e.g.][]{rochester1976secular}. 
In no-slip spherical geometries, it is given by \citep[see formula 35 in][]{rochester1976secular}
\begin{equation}
    \boldsymbol{\Gamma}_\nu \propto E^{1/2} \left [ \mathrm{Re}(\tau_\text{so}) \, \boldsymbol{\omega}_\perp - \mathrm{Im}(\tau_\text{so}) \, \boldsymbol{1}_z \times \boldsymbol{\omega}_\perp + \tau_\text{sup} \, \omega_z \boldsymbol{1}_z \right ]
    \label{eq:torqueNS}
\end{equation}
at the leading order in $E$ (assuming $\boldsymbol{\Omega} = \boldsymbol{1}_z$), where $\boldsymbol{\omega} = \boldsymbol{\omega}_\perp + \omega_z \boldsymbol{1}_z = (\omega_x, \omega_y, \omega_z)^\top$ is the uniform vorticity of the forced flow.
Note that similar expressions have been later rediscovered for the particular case of precession as viewed in the precession frame \citep[e.g.][]{noir2003experimental,noir2013precession}. 
Formula (\ref{eq:torqueNS}) clearly shows that the equatorial components $\boldsymbol{L} \times \boldsymbol{1}_z$ are damped by viscosity when $\mathrm{Re}(\tau_\text{so}) \neq 0$ and, similarly, $\tau_\text{sup}\neq0$ (which is a real number for this mode) ensures that the axial angular momentum $\boldsymbol{L} \boldsymbol{\cdot} \boldsymbol{1}_z$ is affected by viscosity.
Since $\mathrm{Re}(\tau_\text{so}) \neq 0$ and $\tau_\text{sup}\neq0$ in no-slip spheres and ellipsoids, we have $\boldsymbol{\Gamma}_\nu \neq \boldsymbol{0}$ from formula (\ref{eq:torqueNS}) such that the angular momentum is affected by viscosity on long time scales for NS-BC.  

Similar reasoning can be applied to the stress-free rotating case.
It can be shown that leading-order viscous torque (\ref{eq:viscoustorque2})  depends on the viscous decay rates $[\tau_\text{so},\tau_\text{sup}]$ for the SF-BC (not given here, since it vainly makes the expression more complex because a full description of the viscous cross-interactions between  $\boldsymbol{Q}_\text{so}$ and $\boldsymbol{Q}_\text{sup}$ is required contrary to the no-slip case). 
We can thus get physical insight into formula (\ref{eq:viscoustorque2}) by computing the viscous decay rates for SF-BC. 
To do so, we expand the velocity as $\boldsymbol{Q}_k + E^{1/2} \widetilde{\boldsymbol{Q}}_k$ \citep{rieutord1992ekman}, where $\widetilde{\boldsymbol{Q}}_k$ is the boundary-layer flow such that $\boldsymbol{Q}_k + E^{1/2} \widetilde{\boldsymbol{Q}}_k$ satisfies SF-BC (\ref{eq:BCSF}).
The viscous decay rate for SF-BC is then given at the leading order in $E$ by \citep[e.g.][]{liao2001viscous}
\begin{equation}
    \tau_k \int_V |\boldsymbol{Q}_k|^2 \, \mathrm{d} V = E \int_V \boldsymbol{Q}_k^\dagger \boldsymbol{\cdot} \nabla^2 (\boldsymbol{Q}_k + E^{1/2} \widetilde{\boldsymbol{Q}}_k) \, \mathrm{d}V.
    \label{eq:tauSF1}
\end{equation}
Contrary to the no-slip case \citep[for which the boundary-layer flow is of the same order of magnitude as the inviscid flow, e.g.][]{greenspan1968theory}, an explicit solution of $\widetilde{\boldsymbol{Q}}_k$ for SF-BC is not required to estimate $\tau_k $ in equation (\ref{eq:tauSF1}).
Indeed, the representative volume-averaged viscous decay rate of all the eigenmodes is given at leading order in $E$ for our SF-BC by \citep[e.g.][]{rieutord1997ekman}
\begin{equation}
    \tau_k \int_V |\boldsymbol{Q}_k|^2 \, \mathrm{d} V = - 2 E \int_V \boldsymbol{\epsilon} (\boldsymbol{Q}_k) : \boldsymbol{\epsilon} (\boldsymbol{Q}_k^\dagger) \, \mathrm{d} V.
    \label{eq:decaySFVol}
\end{equation}
Expression (\ref{eq:decaySFVol}) generalises formula (3.14) in \citet{liao2001viscous}, which is only valid for spheres (see Appendix \ref{appendix:viscous}), to triaxial ellipsoids. 
Since the right-hand side of equation (\ref{eq:tauSF1}) is real, we have $\tau_k \leq 0$ for SF-BC.
Consequently, there is no viscous correction of the inviscid eigenfrequency $\lambda_k$ at the leading order in $E$ for SF-BC \citep[as initially reported in][]{liao2001viscous}. 
Formula (\ref{eq:decaySFVol}) is illustrated in figure \ref{fig:ReGImG} for a particular configuration.
We recover from the formula that $\tau_\text{so}=\tau_\text{sup}=0$ in spherical geometries (since $\boldsymbol{Q}_\text{so}$ and $\boldsymbol{Q}_\text{sup}$ are exact solid-body rotations in spheres), which agrees with the fact that $\boldsymbol{\Gamma}_\nu = \boldsymbol{0}$ in spheres \citep[e.g.][]{jones2011anelastic}. 

Explicit expressions of $\tau_\text{so}$ and $\tau_\text{sup}$ can be obtained for the uniform-vorticity modes in ellipsoids, because the eigenvectors $[\boldsymbol{\omega}_\text{so},\boldsymbol{\omega}_\text{sup}]$ of matrix (\ref{eq:matrixGP1}) can be analytically obtained. 
The analytical formula of $\tau_\text{so}$, which is too lengthy to be given here, shows that $\tau_\text{so} \neq 0$ in every non-spherical geometry. 
The mathematical reason is that the spin-over mode $\boldsymbol{Q}_\text{so}$ is no longer a solid-body rotation in ellipsoids (i.e. $\boldsymbol{\epsilon} (\boldsymbol{Q}_k)$ is non-zero for the spin-over mode in ellipsoids).
Thus, from a physical viewpoint, a non-zero boundary-layer flow $\widetilde{\boldsymbol{Q}}_\text{so}$ is required to match the SF-BC within a thin Ekman boundary layer. 
Since the spin-over mode is damped by viscosity in ellipsoids, the equatorial angular momentum $\boldsymbol{L} \times \boldsymbol{1}_\Omega$ is affected by viscosity on long time scales (even in axisymmetric ellipsoids). 
After little algebra, the decay rate $\tau_\text{sup}$ is explicitly given by
\begin{equation}
    \frac{\tau_\text{sup}}{E} \int_V |\boldsymbol{Q}_\text{sup}|^2 \, \mathrm{d}V = -\frac{16 \pi}{3}  abc \left [ \Omega_x^2 (b^2-c^2)^2 + \Omega_y^2 (a^2-c^2)^2 + \Omega_z^2 (a^2-b^2)^2 \right ],
    \label{eq:tausup}
\end{equation}
where the axial geostrophic mode is $\boldsymbol{Q}_\text{sup} = \omega_{\text{sup},x} \, \boldsymbol{e}_x + \omega_{\text{sup},y} \, \boldsymbol{e}_y + \omega_{\text{sup},z} \, \boldsymbol{e}_z$ with $\omega_{\text{sup},x} = \Omega_x (b^2 + c^2)$, $\omega_{\text{sup},y} = \Omega_y (a^2 + c^2)$ and $\omega_{\text{sup},z} = \Omega_z (a^2 + b^2)$. 
Formula (\ref{eq:tausup}) shows that $\tau_\text{sup} \neq 0$ when $a \neq b \neq c$, illustrating that the axial geostrophic mode is damped by viscosity in triaxial geometries. 
Therefore, the physical reason why $\boldsymbol{\Gamma}_\nu \neq \boldsymbol{0}$ in triaxial stress-free ellipsoids is that the spin-over and geostrophic modes are damped by viscosity (as evidenced by the non-zero decay rates $\tau_\text{so} \neq 0$ and $\tau_\text{sup} \neq 0$ in such geometries). 
Moreover, formula (\ref{eq:tausup}) shows that $\tau_\text{sup}=0$ when $\boldsymbol{\Omega}$ is an axis of revolution of the geometry (i.e. when $\boldsymbol{\Omega} \propto \boldsymbol{1}_x$ if $b=c$, $\boldsymbol{\Omega} \propto \boldsymbol{1}_y$ if $a=c$, or $\boldsymbol{\Omega} \propto \boldsymbol{1}_z$ if $a=b$). 
The axial geostrophic mode is thus unaffected by viscous dissipation, which explains why the long-term evolution of $\boldsymbol{L} \boldsymbol{\cdot} \boldsymbol{1}_\Omega$ is physically unconstrained in such pathological configurations. 
This was the situation previously considered for precession-driven flows in spheroids \citep{lorenzani2003inertial,wu2009dynamo,guermond2013remarks}.
Yet, the conclusion is not valid for every axisymmetric geometry with global rotation. 
Indeed, we have $\tau_\text{sup}\neq 0$ in axisymmetric geometries if $\boldsymbol{\Omega}$ is not the revolution symmetry axis (such that $\boldsymbol{L} \boldsymbol{\cdot} \boldsymbol{1}_\Omega$ will be damped by viscosity).

The BLT of Coriolis eigenmodes has thus explained why the long-term angular momentum evolution is damped by viscosity in triaxial geometries, but also in axisymmetric ellipsoids if the mean rotation axis $\boldsymbol{\Omega}$ is not the revolution symmetry axis.

\subsection{Resonance conditions for mechanical forcings}
\begin{figure}
    \centering
    \begin{tabular}{cc}
        \includegraphics[width=0.49\textwidth]{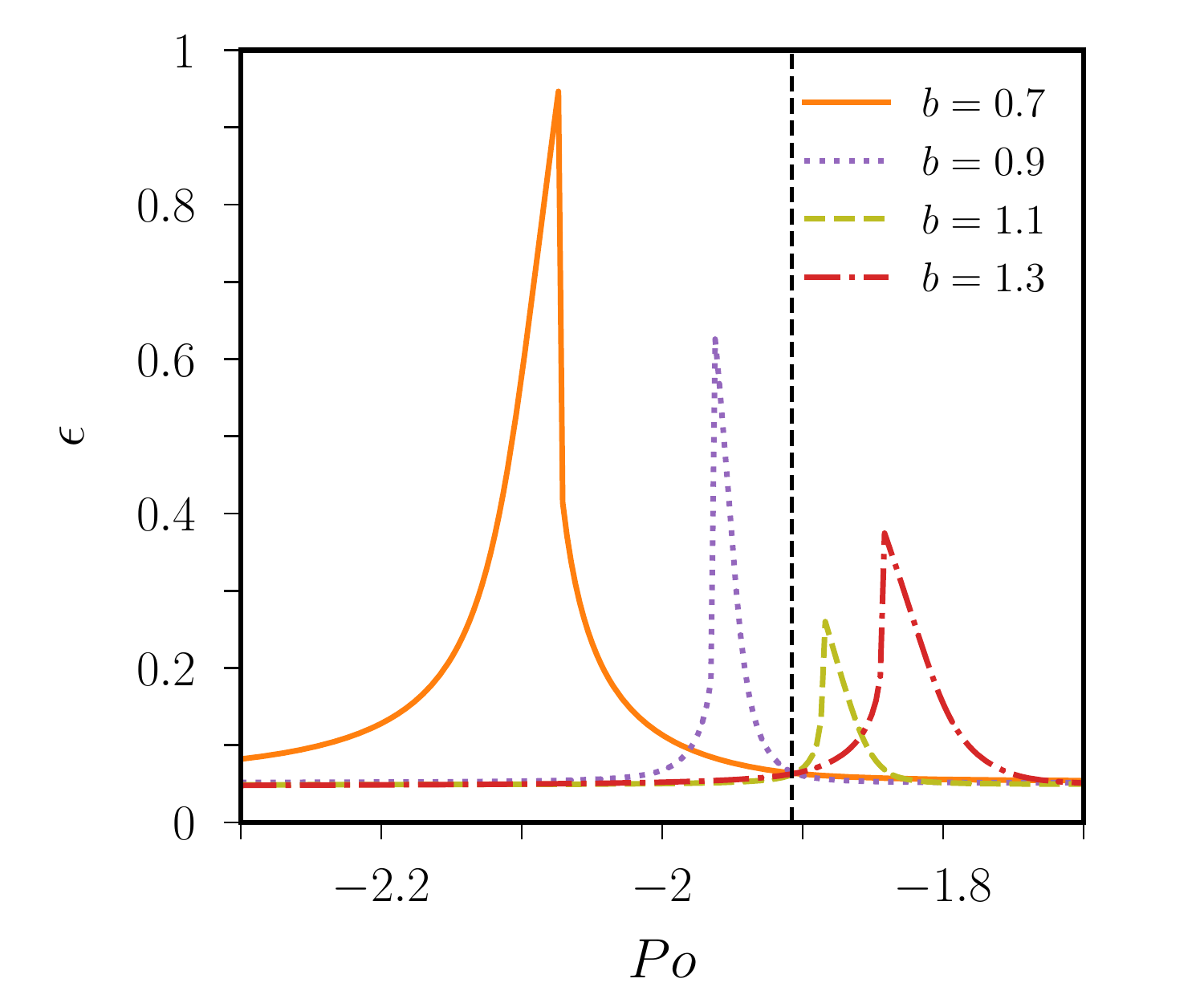} &
        \includegraphics[width=0.49\textwidth]{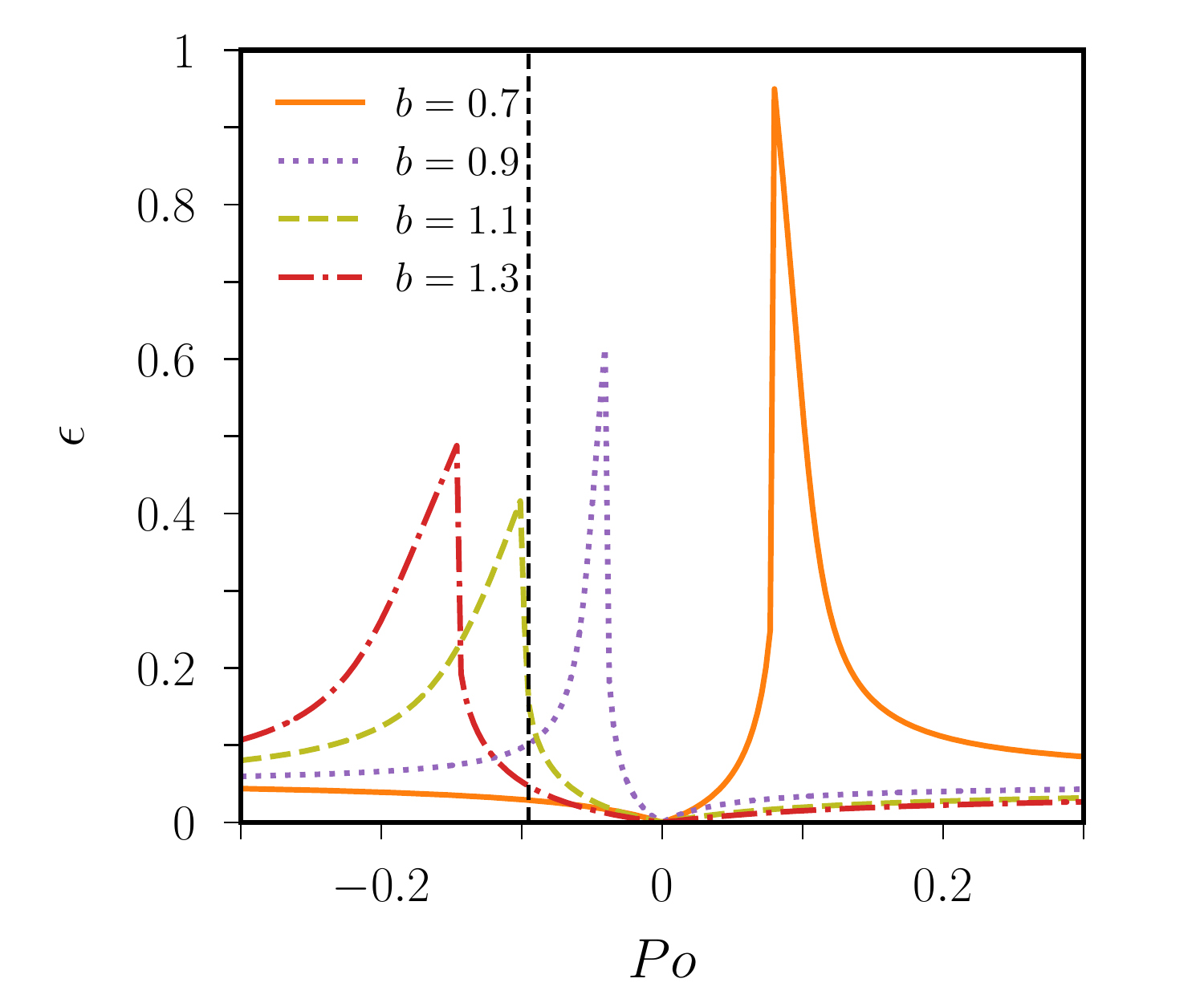} \\
        (a) $ Po^{-}$ & (b) $Po^{+}$ \\
    \end{tabular}
    \caption{Double resonance at $Po^\pm$ of the forced precession-driven flow in ellipsoids for SF-BC with $a=1$ and $c=0.9$ (values of $b$ given in the legend). Time-averaged differential rotation $\epsilon = \overline{|\boldsymbol{\omega}|}$ of numerical solutions of equation (\ref{eq:ODEGP1}) at $E=10^{-3}$ and small precession angle $\alpha = 3^\circ$. Vertical dashed lines show $Po^{\pm}$ predicted by equation (\ref{eq:ResonanceBF1}) at $b=1$.}
    \label{fig:resonance}
\end{figure}

A key property of the primary uniform-vorticity flow is its ability to enter in direct resonance with the precession forcing (as evidenced by the divergent amplitude of the asymptotic solution). 
A direct resonance requires a close spatial and temporal matching between the Poincar\'e force and the flow response \citep[e.g.][]{greenspan1968theory}. 
The spatial matching is ensured by the fact that both the Poincar\'e force and the forced uniform-vorticity flow are linear in the Cartesian coordinates. 
Heuristically, the temporal resonance condition requires that the frequency $\omega_p$ of the forcing (for monochromatic forcings) must be equal (or close) to the angular frequency $f$ of the forced flow in the mantle frame, which gives $f = \pm \, \omega_p$. 
The latter condition generally predicts the existence of two resonances for mechanically driven flows in ellipsoids (if the spatial resonance conditions are satisfied). 
A quick inspection of equation (\ref{eq:ODEGP1}) shows that the uniform-vorticity dynamics roughly corresponds to that of a harmonic oscillator driven by the Poincar\'e force in the inviscid regime $E=0$. 
Consequently, direct resonances occur when the forced flow corresponds to a free oscillatory eigenmode of the unforced system, namely the spin-over mode $\boldsymbol{Q}_\text{so}$ such that $f \propto \lambda_\text{so}$ (up to a normalisation prefactor). 
For this reason, longitudinal librations (which only directly excite the zero-frequency geostrophic mode) do not exhibit any inviscid resonance in spheres \citep[e.g.][]{zhang2013non} or ellipsoids. 
On the contrary, latitudinal librations can trigger the spin-over mode and the corresponding forced laminar flow exhibits two inviscid resonances occurring at $ \lambda_\text{so} = \pm \, \omega_p$ in non-spherical geometries \citep{zhang2012asymptotic,vantieghem2015latitudinal}, where $\omega_p$ is the libration angular frequency. Similarly, a second resonance has already been found for the interaction between tides and precession in triaxial ellipsoids \citep{cebron2010tilt}. 
A second resonance for pure precession is thus also expected in ellipsoids from simple theoretical arguments. 
Assuming that the forced uniform-vorticity flow is oscillating in the mantle frame at the effective angular frequency $f \simeq \left [1 + P_z \right ] \lambda_\text{so}$ when $|P_x| \ll 1$, the temporal resonance condition predicts two direct resonances for precession at the resonant Poincar\'e numbers $Po^\pm$ given by
\begin{equation}
	1 + Po^{\pm} \, \cos (\alpha)  = \pm 1/\lambda_\text{so},
	\label{eq:ResonanceBF1}
\end{equation}
where $\lambda_\text{so} = 2 ab/\sqrt{(a^2+c^2)(b^2+c^2)}$ is here the eigenfrequency of the spin-over mode in equation (\ref{eq:matrixGP1}) with $\boldsymbol{\Omega}=\boldsymbol{1}_z$ \citep[see also formula 3.21 in][]{vantieghem2014inertial}.
The above condition is exactly the resonance condition of asymptotic solution (\ref{eq:PrecTriax1}).
The two resonances at $[Po^-, Po^+]$ are thus robust features of precession-driven flows, but it remains to elucidate why the second resonance at $Po^-$ has not been reported before.

\begin{figure}
    \centering
    \begin{tabular}{cc}
        \includegraphics[width=0.49\textwidth]{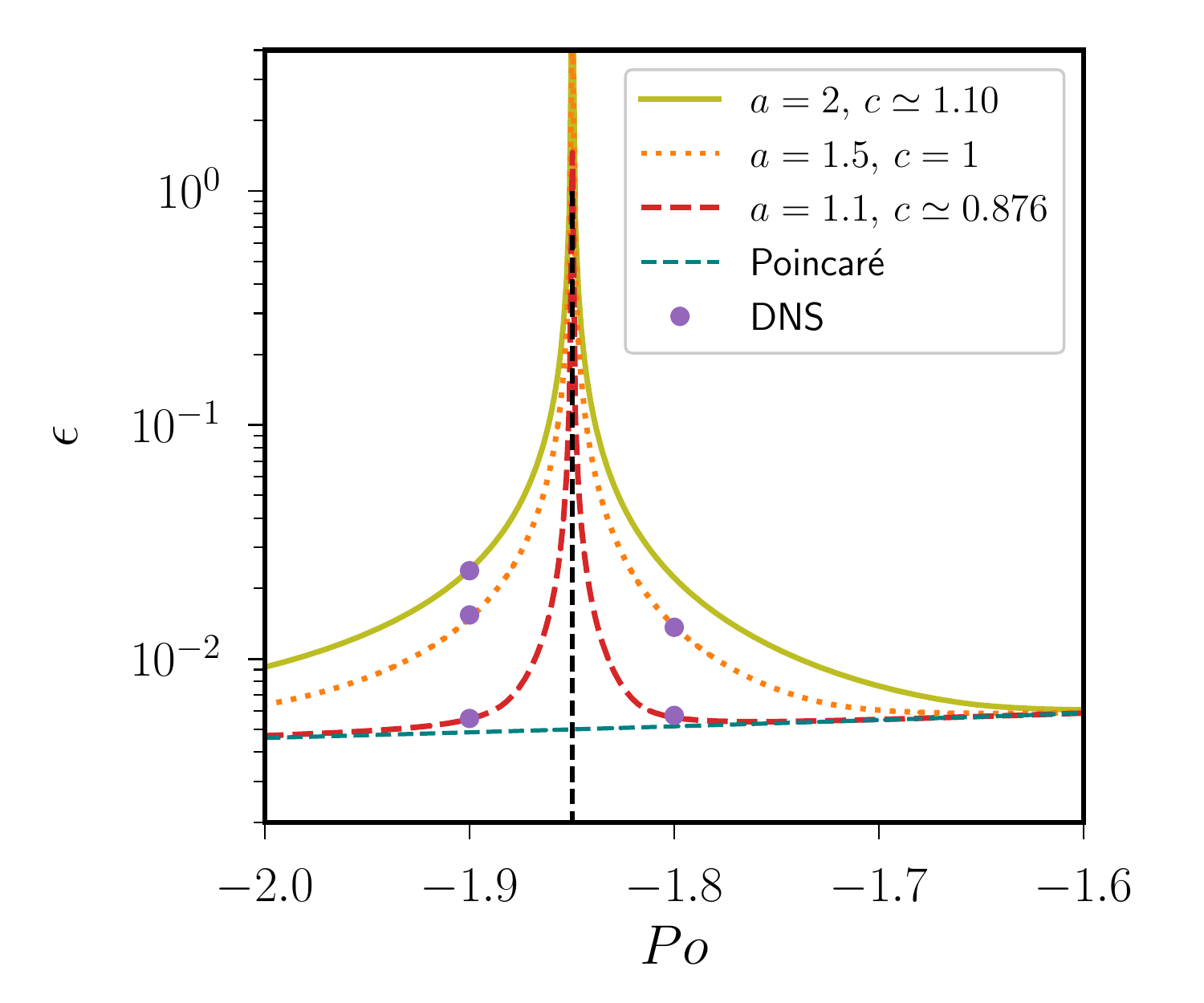} &
        \includegraphics[width=0.49\textwidth]{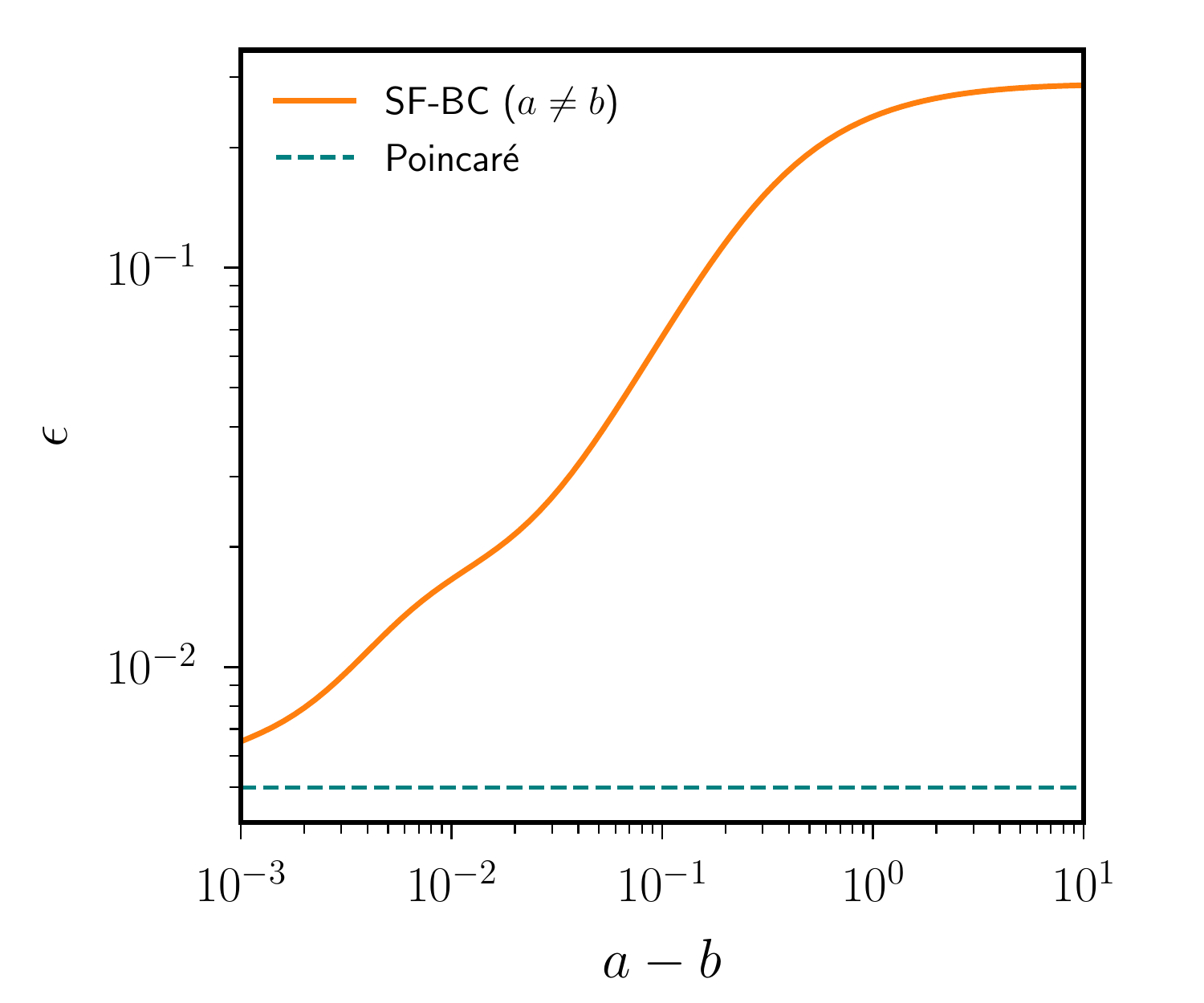} \\
        (a) & (b)  \\
    \end{tabular}
    \caption{Behaviour of the forced flow near the second resonance $Po^{-}$ in stress-free ellipsoids with $b=1$ and $P_x=10^{-2}$. The resonant value is fixed at the value $Po^-$ obtained with $a=1.5$ and $c=1$ (as in figure \ref{fig:benchmark}). To maintain a fixed resonance when $a$ is varied, the polar axis is given by $c= 0.5 [-2 a^2 - 2 + 2 \sqrt{-32 a^2 \, \Delta^{1/2} + 1 + a^4 + 2 \, (8 \Delta + 7) a^2}]^{1/2}$ with $\Delta = (Po^{-} - P_x)(Po^{-} + P_x)$. In the two panels, the dashed teal line shows the expected inviscid value from Poincar\'e solution (\ref{eq:Poincare}) for $a=b=1$. (a) Comparison between DNS at $E=5\times10^{-4}$ and asymptotic solution (\ref{eq:PrecTriax1}). (b) Numerical solutions of equation (\ref{eq:ODEGP1}) for SF-BC at $Po=Po^-$ and $E=10^{-3}$.}
    \label{fig:vanishing}
\end{figure}

We have numerically solved equation (\ref{eq:ODEGP1}) in time to explore the behaviour of the solutions near the double resonances in figure \ref{fig:resonance}. 
The two resonances at $Po^\pm$ are continuously shifted when $b$ is varied and, at $b=1$, the resonant value $Po^-$ differs from $Po^+$ as observed in panel (b).
This directly results from condition (\ref{eq:ResonanceBF1}), which predicts that the two resonances are linked by $[Po^+ + Po^-] \cos(\alpha) = -2$.
This clearly shows that the two direct resonances do not merge together in ellipsoids. 
We further explore the behaviour near $Po^-$ in figure \ref{fig:vanishing}.  
We have fixed the resonant value $Po^-$ at its value given in figure \ref{fig:benchmark} for $a=1.5$ and $b=c=1$ and, then, adjusted the polar axis $c$ to maintain the resonance at $Po^-$ for different values of $a$. 
We observe that the width of the resonance peak decreases when $a\to b$ (panel a).
This is a purely inviscid feature of the asymptotic solution, which is recovered in the DNS.  
The particular case $a=b$ is not formally defined for SF-BC, but it can be approached by decreasing $a-b$ (panel b). 
The amplitude of the stress-free solution at $Po=Po^-$ is limited by the viscosity and approaches, when $a \to b$, the inviscid Poincar\'e solution for $a=b$.
The differential rotation $\epsilon$ of the inviscid Poincar\'e solution is given by \citep[assuming $\omega_z = 1$, see Appendix B in][]{wu2011high}
\begin{equation}
    \epsilon = \left | \frac{P_x (2 + \eta)}{\eta+2(1+\eta)P_z}\right |,
    \label{eq:Poincare}
\end{equation}
which is non-divergent when $Po=Po^-$.
This agrees with a lengthy mathematical analysis of the behaviour near the inviscid resonances (not given here), which shows that the second inviscid resonance at $Po^-$ disappears in spheroids with $a=b$ contrary to the other resonance at $Po^+$ \citep[e.g.][]{busse1968steady,noir2013precession}.

We have thus understood why precession-driven flows are subject to two inviscid resonances in triaxial ellipsoids, which occur at the resonant Poincar\'e numbers $Po^\pm$ given by equation (\ref{eq:ResonanceBF1}) when $|P_x| \ll 1$. 
Since the two resonances are inviscid features of the forced flow in ellipsoids, they exist for both SF-BC and NS-BC. 
The second resonance actually disappears in spheroidal geometries $a=b$ (i.e. its amplitude is vanishing), which explains why previous works in spheroids have not observed it \citep[e.g.][]{cebron2015bistable,nobili2021hysteresis}.
Previous studies in triaxial geometries \citep[e.g.][]{noir2013precession,burmann2022experimental} have also overlooked it, because it usually occurs at $|Po^-| \gg |Po^+|$.

\subsection{Implications for DNS}
\label{subsec:DNS}
\begin{figure}
    \centering
    \begin{tabular}{cc}
    \begin{tabular}{c}
        \includegraphics[width=0.49\textwidth]{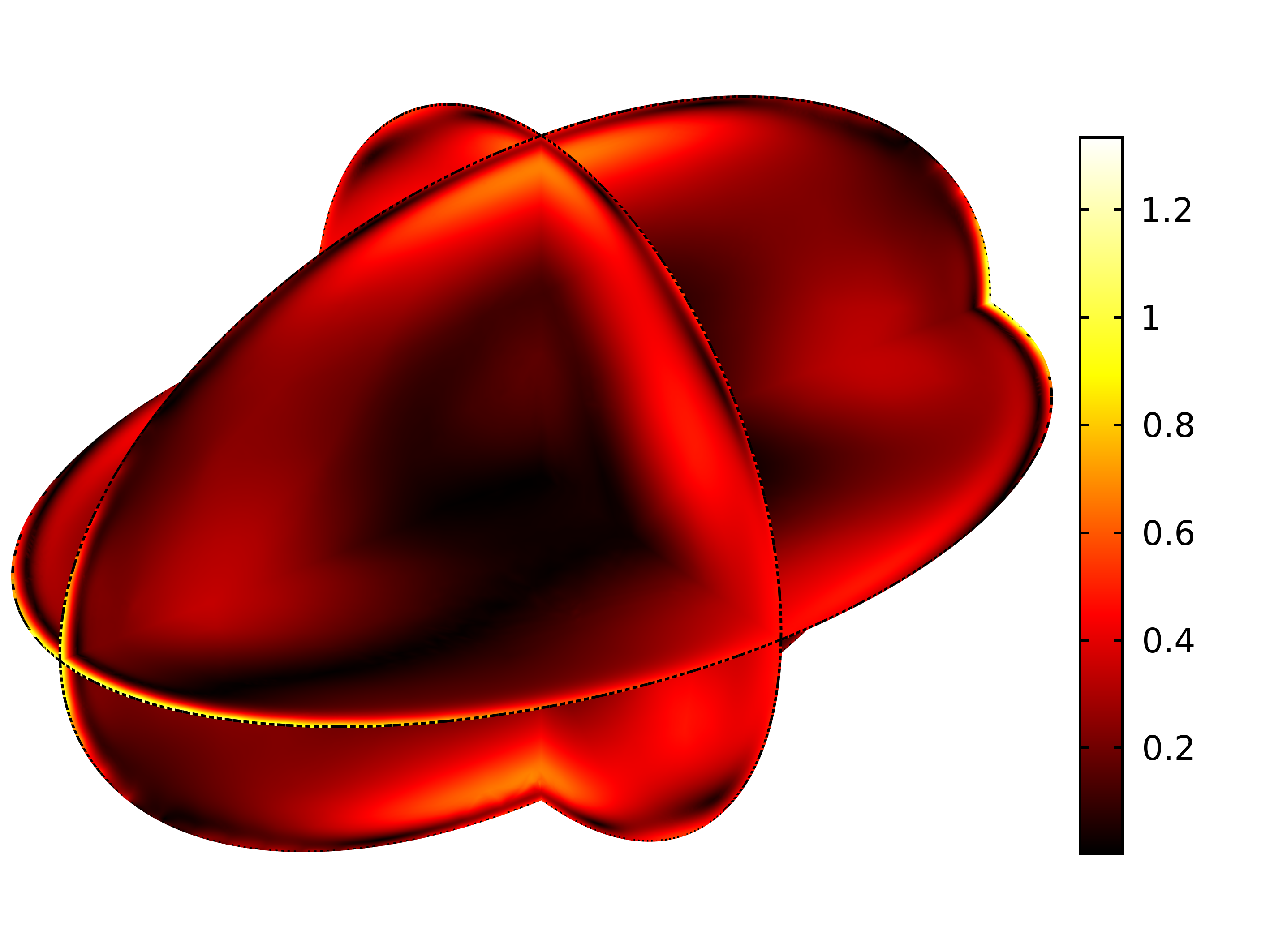}\\
    \end{tabular} &
    \begin{tabular}{c}
        \includegraphics[width=0.49\textwidth]{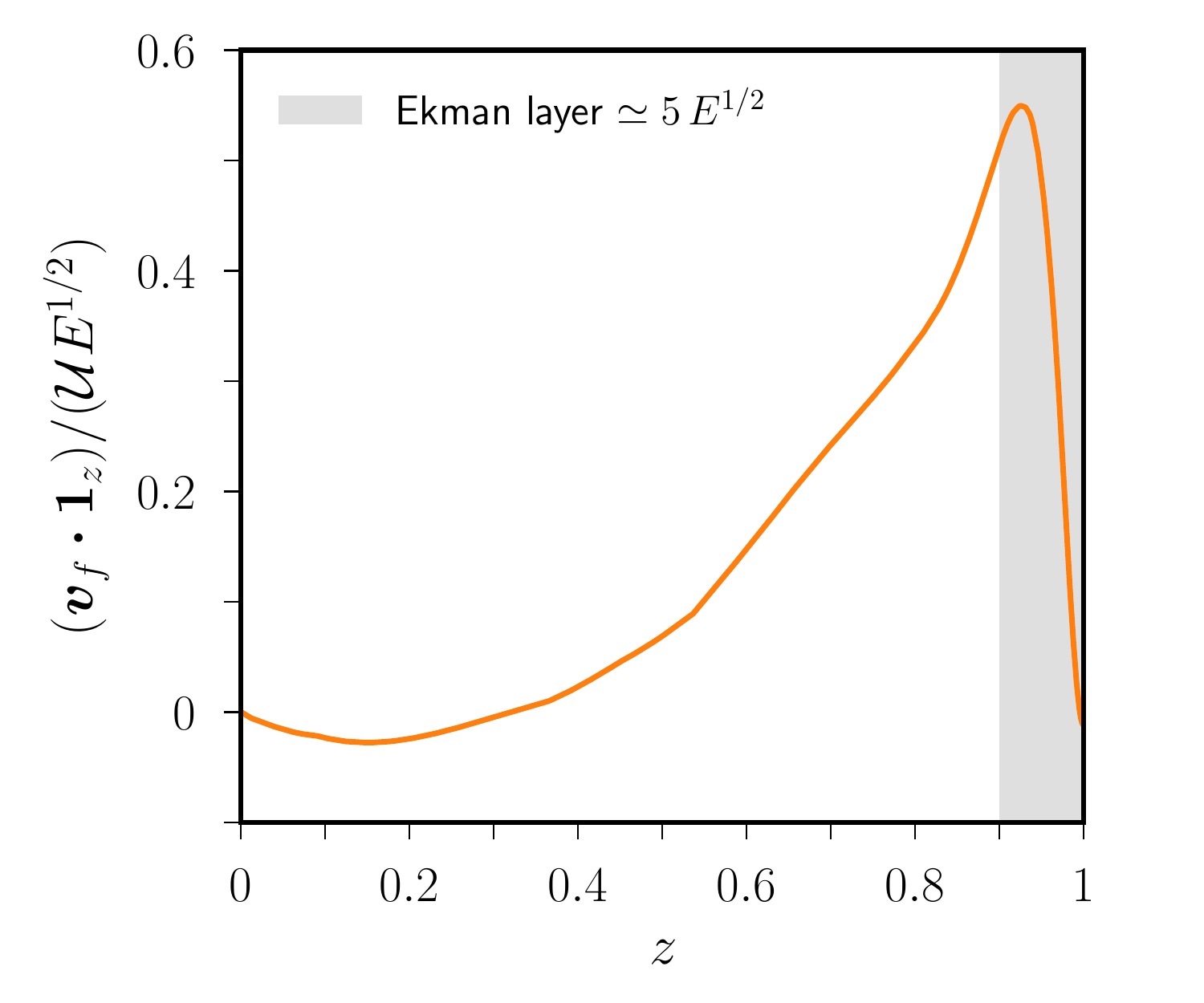} \\
    \end{tabular} \\
    (a) & (b) \\
    \end{tabular}
    \caption{DNS of precession-driven flow with SF-BC at $Po=-1.8$, $P_x = Po \sin (\alpha) = 10^{-2}$ and $E=5 \times 10^{-4}$. 
    Axisymmetric geometry $a=1.5$ and $b=c=1$. 
    Normalised velocity $\boldsymbol{v}_f/(\mathcal{U} E^{1/2})$, as defined in expansion (\ref{eq:expansionBuffett}), at time $t=39530$ where $\mathcal{U}=0.0129$ is the maximum of $|\boldsymbol{v}_f|$.
   (a) Three-dimensional rendering of the velocity magnitude using a linear scale. 
   (b) Axial velocity component as a function of $z$ along the $c$-axis.}
    \label{fig:Ekman}
\end{figure}

We have shown that the long-term evolution of angular momentum is affected by viscosity, due to the existence of an Ekman boundary layer in rapidly rotating ellipsoids.
The uniform-vorticity elements carrying angular momentum in expansion (\ref{eq:expansionBuffett}) do not indeed satisfy the SF-BC in triaxial geometries.
Thus, they are associated with an Ekman boundary layer to match the boundary conditions. 
This is a noticeable difference with the more usual spherical geometry, in which $\boldsymbol{\Gamma}_\nu = \boldsymbol{0}$ \citep[e.g.][]{jones2011anelastic}. 
The Ekman boundary layer in ellipsoids is clearly observed in figure \ref{fig:Ekman}. 
Its typical thickness is still $\mathcal{O}(E^{1/2})$ but, contrary to the case of NS-BC, the amplitude of the boundary-layer flow is $\mathcal{O}(E^{1/2})$ smaller than the bulk flow amplitude \citep[in agreement with][]{rieutord1992ekman}. 

This could have implications for numerical studies using stress-free boundaries. 
A numerical strategy has to be employed to ensure the conservation of angular momentum in spherical codes \citep[e.g.][]{jones2011anelastic}.
This is no longer necessary in triaxial ellipsoids since $\boldsymbol{\Gamma}_\nu \neq \boldsymbol{0}$ \citep[albeit such a strategy may be considered to ensure the conservation of the axial angular momentum if the mean rotation axis is an axis of revolution symmetry, as proposed in][]{guermond2013remarks}. 
However, for the moderate values of the Ekman number achievable in DNS, the flow within the Ekman layer will modify the value of the viscous torque (which pilots the long-term evolution of angular momentum). 
Indeed, instead of using expression (\ref{eq:viscoustorque1}), the viscous torque is usually computed with the surface integral $\boldsymbol{\Gamma}_\nu = 2 \, E \int_S \boldsymbol{r} \times ( \nabla \boldsymbol{\cdot} \boldsymbol{\epsilon} ) \, \mathrm{d} S$ as
\begin{equation}
    \boldsymbol{\Gamma}_\nu = 2 \, E \int_S \boldsymbol{r} \times \boldsymbol{\mathcal{T}} \, \mathrm{d} S = 2 \, E \int_S \boldsymbol{r} \times \left [ (\boldsymbol{\mathcal{T}} \boldsymbol{\cdot} \boldsymbol{1}_n) \, \boldsymbol{1}_n \right ]  \mathrm{d} S,
    \label{eq:ViscousS}
\end{equation}
in which we have used formula (9) in \cite{rochester1962geomagnetic} for a symmetric tensor to obtain the first equality and, then, have written the surface traction as $\boldsymbol{\mathcal{T}} = (\boldsymbol{\mathcal{T}} \boldsymbol{\cdot} \boldsymbol{1}_n) \, \boldsymbol{1}_n - \boldsymbol{1}_n \times (\boldsymbol{1}_n \times \boldsymbol{\mathcal{T}}) = (\boldsymbol{\mathcal{T}} \boldsymbol{\cdot} \boldsymbol{1}_n) \, \boldsymbol{1}_n$ on the boundary for SF-BC (\ref{eq:BCSF}). 
Formula (\ref{eq:ViscousS}) shows that the normal component of the surface traction, which is non-zero in the presence of an Ekman layer in stress-free ellipsoids, contributes to the viscous torque. 
Hence, numerical and local approximations of SF-BC (\ref{eq:BCSF}) have no reasons to yield a vanishing torque component in formula (\ref{eq:ViscousS}) for axisymmetric ellipsoids if the boundary layer is not sufficiently resolved (as observed in some DNS, not shown). 
Using a refined boundary-layer mesh may thus be required to properly describe the Ekman layer in ellipsoids and ensure sufficient torque accuracy (which can be used to check the numerical convergence).

\subsection{Scaling laws}
\label{subsec:planets}
Despite the existence of a thin Ekman layer, we believe that adopting SF-BC in global simulations is useful to probe bulk mechanisms that can be hampered by viscous effects when NS-BC are employed. 
The case of precession is illuminating in this respect.
Indeed, the laminar precession-driven flow can be destabilised by several hydrodynamic instabilities in no-slip ellipsoids, such as the inertial (topographic) instabilities outlined in \S\ref{subsec:instab} and the conical-shear instability (CSI).
The former are due to the ellipticity of the boundary and survive in the inviscid regime $E = 0$.
On the contrary, the CSI is a parametric instability existing because of the viscous conical shear layers spawned from the Ekman layer at the critical latitudes \citep{lin2015shear}. 
In addition, precession also often triggers boundary-layer instabilities within the Ekman layer for NS-BC \citep[e.g.][]{lorenzani2001fluid,cebron2019precessing,buffett2021conditions}. 
A comprehensive study of these instabilities deserves further work, but we can estimate their relevance as follows. 
As outlined in \S\ref{subsec:instab}, the typical inviscid growth rate of the precession-driven inertial instabilities is given by formula (\ref{eq:sigmaTopo}) for the large-scale modes.
For the CSI, the growth rate in full spheres and ellipsoids is given by \citep{lin2015shear,horimoto2020conical}
\begin{equation}
    \sigma_\text{CSI} = \mathcal{O}(\epsilon E^{1/5}).
    \label{eq:sigmaCSI}
\end{equation}
Quantitatively, a necessary condition for the existence of the two instabilities is that growth rates (\ref{eq:sigmaTopo}) and (\ref{eq:sigmaCSI}) are larger than the viscous damping. 
For the NS-BC, this damping is mainly due to the Ekman layer and its amplitude is of the order $\mathcal{O}(E^{1/2})$ \citep{greenspan1968theory}. 
Actually, it appears that large-scale inertial instabilities are difficult to obtain for the moderately small values of the Ekman number usually considered in experiments or DNS (as outlined in figure \ref{fig:floquet}). 

\begin{figure}
    \centering
    \begin{tabular}{cc}
        \includegraphics[width=0.49\textwidth]{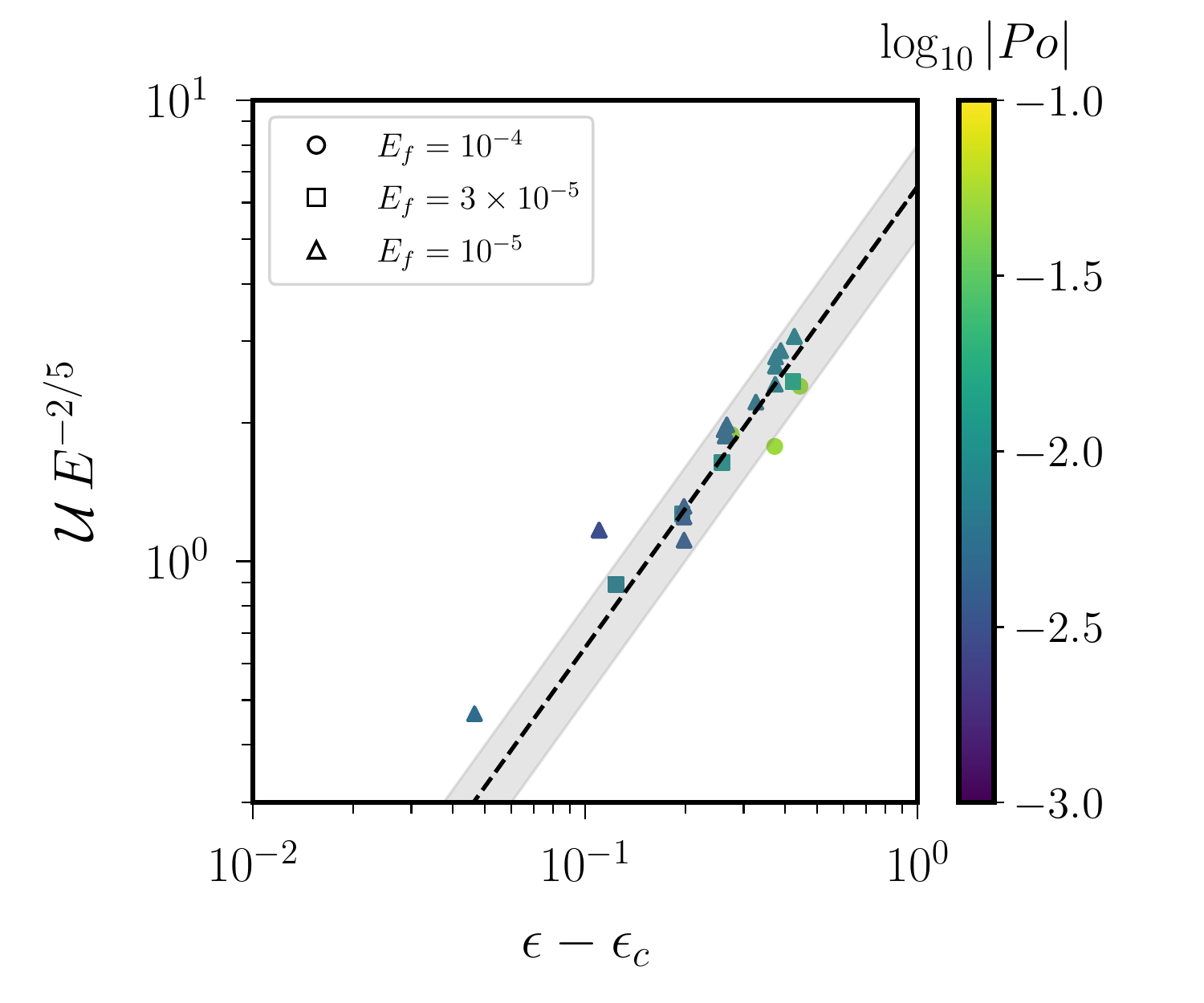} &
        \includegraphics[width=0.49\textwidth]{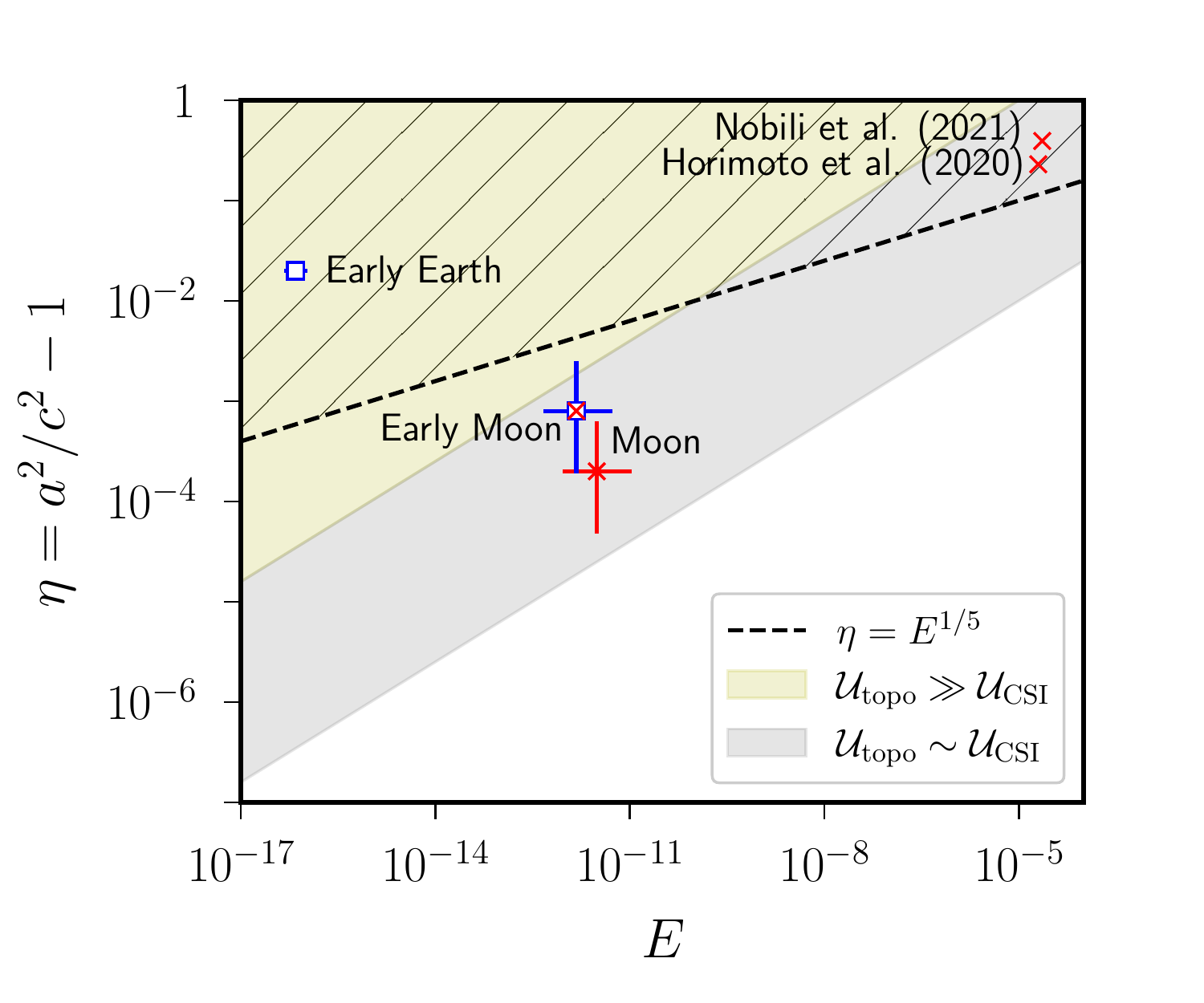} \\
        (a) & (b) \\
    \end{tabular}
    \caption{(a) Comparison between scaling law (\ref{eq:CSIsat}) and DNS for $|Po| < 0.1$ in no-slip full spheres from the database of figure 7 in \citet{cebron2019precessing}, with $E_f = E/|1+Po| \simeq E $ when $|Po| \ll 1$. 
    Colour bar indicates $\log_{10} |Po|$.
    Grey area shows the scaling law $\mathcal{U} \, E^{-2/5} = (6.5 \pm 1.5) \, (\epsilon-\epsilon_c)$ and dashed line $\mathcal{U} \, E^{-2/5} = 6.5 \, (\epsilon-\epsilon_c)$, where $\epsilon_c \approx 7 E_f^{3/10}$ is an estimate of the onset value \citep[see equation 17 in][]{cebron2019precessing}. 
    (b) Competition between the inertial instabilities and the CSI in precessing ellipsoids. Empty blue squares $\color{blue}{\square}$ show conditions for which inertial (topographic) instabilities are expected, and red crosses $\color{red}{\times}$ indicate where the CSI is expected or observed. 
    Grey area shows $\eta \propto E^{2/5}$ with a (unknown) prefactor chosen in the range $[1,100]$, in which we expect $\mathcal{U}_{\text{topo}} \sim \mathcal{U}_{\text{CSI}}$ when both instabilities exist.
    Hatched area is the region where $\sigma_\mathrm{topo} \gtrsim \sigma_\mathrm{CSI}$ (if the two instabilities coexist).
    White area is the region where  $\mathcal{U}_{\text{topo}} \ll \mathcal{U}_{\text{CSI}}$. 
    Estimates for the early Moon and Earth taken from Appendix \ref{appendix:planets} with $\eta \approx 2 f$.}
    \label{fig:csi}
\end{figure}

A linear analysis is, however, not sufficient to determine the physical relevance of these instabilities.
In particular, scaling laws are worth finding to estimate the strength of the precession-driven flows driven by such instabilities.  
Indeed, the inertial instabilities have presumably a saturation amplitude almost independent of the Ekman number \citep[as found for the turbulence driven by tidal instabilities, e.g.][]{grannan2017tidally}, whereas the CSI amplitude could decrease when $E \to 0$ as the instability results from viscous effects. 
A rigorous description of the nonlinear regimes requires dedicated simulations, but the saturation amplitudes can be crudely estimated using simple order-of-magnitude arguments \citep[which have already proven useful for tidal flows, e.g. in][]{barker2013non,barker2016turbulence}. 
We assume that the flow amplitude $\mathcal{U}$ resulting from the primary instability grows until secondary instabilities, characterised by the growth rate $\sigma_\text{sec}$, become strong enough to prevent further growth of the primary instability.
A saturated turbulent regime would then be obtained when $\mathcal{U} \sim \sigma_\text{sec} \ell$, where $\ell$ is a characteristic length scale of the primary unstable flow.
The nonlinear saturation of the inertial (topographic) instabilities would thus be given by (in dimensionless units)
\begin{equation}
    \mathcal{U}_\text{topo} = \mathcal{O}(\epsilon \eta)
    \label{eq:Toposat}
\end{equation}
with $\ell \sim 1$ for a large-scale instability.
A good agreement with the above scaling law has been found using DNS in shearing periodic boxes \citep{barker2016turbulence}, but the scaling law might be different for short-wavelength instabilities with $\ell \ll 1$.
Using the same reasoning for the CSI, the relevant length scale is likely the width of the critical shear layer $\ell \sim E^{1/5}$ \citep{lin2015shear}.
Assuming that the CSI is limited by secondary CSI within the critical shear layers, we obtain the (dimensionless) scaling law
\begin{equation}
    \mathcal{U}_\text{CSI} = \mathcal{O}(\epsilon E^{2/5}).
    \label{eq:CSIsat}
\end{equation}
We compare in figure \ref{fig:csi}(a) the above scaling law with previously published DNS in no-slip full spheres \citep{lin2015shear,cebron2019precessing}.
Considering the full sphere geometry allows us to discard the possible CSI resulting from the inner boundary (which would give a different scaling). 
However, even in the full sphere, identifying the instability mechanism is difficult due to the competition between the CSI and the boundary-layer instabilities. 
Moreover, due to the non-trivial dependence of the two viscously driven instabilities with the forcing parameters, it is unlikely that a single scaling law could fully describe the entire simulation dataset.
The onset distance is indeed difficult to estimate \citep[e.g. see figure 6 in][]{cebron2019precessing}.
Besides, the simulations may not be in the asymptotic regime $E \ll 1$. 
Despite such uncertainties, a fairly good agreement is found between the DNS and scaling law (\ref{eq:CSIsat}) sufficiently far from the onset. 
This suggests that the CSI was present in the nonlinear regime and that its saturation amplitude obeys scaling law (\ref{eq:CSIsat}) for sufficiently small values of the Ekman number. 

Finally, the comparison between scaling laws (\ref{eq:Toposat}) and (\ref{eq:CSIsat}) shows that the inertial instability would have a larger amplitude than the CSI when $\eta \gg E^{2/5}$ (if the two instabilities were simultaneously triggered). 
The resulting regime diagram is illustrated in figure \ref{fig:csi}(b), using planetary estimates given in Appendix \ref{appendix:planets}. 
Precession-driven inertial instabilities may only have been excited in the primitive liquid cores of the Earth and Moon, whereas the CSI is expected to be present (respectively absent) in the core of the Moon (respectively the Earth) during its whole history \citep{lin2015shear,landeau2022sustaining}. 
In the early Moon, the inertial instabilities may have dominated the CSI in flow amplitude (although the CSI may have had a larger growth rate than the inertial instabilities according to previous formulas, not shown).  
Therefore, the inertial instabilities may actually be more relevant than the CSI for some planetary conditions \citep[although they have not been convincingly observed yet in experiments, e.g.][]{nobili2021hysteresis,burmann2022experimental}.
This could be key for the generation of planetary magnetic fields, as initially postulated for the geodynamo \citep{malkus1968precession}. 
Preliminary estimates of the dynamo capability of the precession-driven instabilities, obtained using (speculative) order-of-magnitude arguments, are given in Appendix \ref{appendix:planets}. 

\section{Conclusion}
\label{sec:conclusion}
    \subsection{Summary}
We have investigated precession-driven flows in stress-free ellipsoids, using asymptotic analysis and targeted DNS.  
We have developed a reduced model for SF-BC to determine the forced uniform-vorticity flows, which carry angular momentum. 
We have shown that angular momentum is affected on long time scales by viscosity in triaxial ellipsoids, but also in axisymmetric geometries if the mean rotation axis is not a revolution symmetry axis. 
This is a noticeable difference from spherical geometries, in which angular momentum is unaffected by viscosity. 
The fundamental reason is that the flows carrying a non-zero angular momentum in ellipsoids are associated with an Ekman boundary layer in rotating ellipsoids. 
From a numerical viewpoint, a boundary-layer mesh may be necessary to get numerical convergence of the angular momentum in rotating ellipsoids. 
We also have obtained the analytical solution of the time-dependent laminar flow forced by precession in the mantle frame, which is valid for planetary parameters and triaxial geometries. 
The comparison with the DNS has shown that, even for moderately small values of the Ekman number, the forced laminar flow in the DNS converges to the asymptotic solution in the vanishing viscosity regime.
Moreover, we have uncovered a second (inviscid) resonance of the forced laminar flow in triaxial ellipsoids.

Then, we have explored the inertial instabilities growing upon the forced laminar flow in the bulk, which survive in the inviscid regime $E=0$. 
We have shown that these instabilities could be more easily observed in stress-free ellipsoids than in no-slip ones (at least for the moderate values $E \gtrsim 10^{-6}$ considered in DNS). 
We have finally proposed scaling laws for the velocity amplitude of the inertial instabilities and of the CSI, which are in good agreement with previous DNS. 
The comparison between the two scaling laws confirms that replacing NS-BC with SF-BC in the mantle frame could be useful to directly probe scenarios of bulk turbulence in the low-viscosity regime (which are of interest for planetary modelling).

    \subsection{Perspectives}
Despite the presence of a thin Ekman boundary layer, we believe that SF-BC are relevant for global models of mechanically driven flows.
The stress-free model could be used to investigate the saturated flows driven by the inertial (topographic) instabilities in precession ellipsoids and, then, their dynamo capability for planetary applications (as outlined in Appendix \ref{appendix:planets}). 
Stress-free models could indeed shed new light on alternative mechanisms giving birth to dynamo fields in planetary interiors.
For instance, the past dynamo of the Moon may have been driven by precession \citep[e.g.][]{dwyer2011long}.
Yet, previous numerical investigations of precession-driven dynamos failed to reproduce large-scale magnetic fields in spherical geometries \citep{cebron2019precessing}.
This could result from the fact that the turbulence was driven in those simulations by viscous flows (e.g. the CSI or boundary-layer instabilities), which may be negligible in amplitude compared with the turbulence driven by the inertial (topographic) instabilities in the early Moon (as discussed in \S\ref{subsec:planets}). 
This hypothesis could be tested in simulations using stress-free ellipsoids. 
Similarly, energetic arguments suggest that the dynamo of the early Earth may have been sustained by tidal flows \citep{landeau2022sustaining}. 
However, the associated fluid dynamics remains to be quantitatively studied to go beyond prior proof-of-concept simulations \citep{reddy2018turbulent,vidal2018magnetic}. 
Precessing stress-free ellipsoids are also relevant for short-period hot Jupiters \citep{barker2016turbulence}, or gaseous planets with a big moon outside the equatorial plane \citep[e.g. the Neptune/Triton pair,][]{wicht2010theory}. 

Finally, SF-BC could also be used to revisit the long-standing problem associated with the generation of geostrophic flows in rotating fluids \citep{greenspan1969non}. 
Nonlinear interactions within the Ekman boundary layers for NS-BC \citep[e.g.][]{busse1968steady,cebron2021mean} or in the bulk through the action of the Reynolds stresses \citep[e.g.][]{zhang2004new,livermore2016comparison}, are usually invoked, but geostrophic flows can also result from bulk turbulence.
However, it remains unclear whether two- or three-dimensional rotating bulk turbulence is established in natural systems \citep[e.g.][]{le2019experimental}. 
This fundamental problem has been attacked in cylindrical or plane-layer geometries \citep[e.g.][]{kerswell1999secondary,brunet2020shortcut,le2020near}. 
Yet, the latter geometries are not directly relevant for planetary modelling, due to the absence of the so-called topographic beta effect that strongly modifies the geostrophic flows in spheres and ellipsoids \citep[e.g.][]{greenspan1968theory}.
We believe that using SF-BC opens the way for new fundamental studies dealing with the interplay between waves and geostrophic flows in global geometries. 

\backsection[Acknowledgements]{We acknowledge the three anonymous referees for their constructive criticisms, which
considerably improved the quality of the manuscript. We also acknowledge the editor, N. Balmforth, for his careful editorial work.}

\backsection[Funding]{This work received funding from the European Research Council (ERC) under the European Union's Horizon 2020 research and innovation programme via the \textsc{theia} project (grant agreement no. 847433). 
ISTerre is part of Labex OSUG@2020 (ANR10 LABX56).}

\backsection[Declaration of interest]{The authors report no conflict of interest.}


\backsection[Author ORCIDs]{\\
\orcidicon J\'er\'emie Vidal \href{https://orcid.org/0000-0002-3654-6633}{https://orcid.org/0000-0002-3654-6633};\\
\orcidicon David C\'ebron  \href{https://orcid.org/0000-0002-3579-8281}{https://orcid.org/0000-0002-3579-8281}.}

\backsection[Author contributions]{The paper is an idea of J.V., who designed the study, conducted the asymptotic theory and developed the bespoke numerical code. 
D.C. conducted the finite-element computations using \textsc{comsol}, and analytically obtained the second-order geostrophic flow. 
Both authors discussed and approved the results presented in the article. 
J.V. drafted the paper, and both authors gave final approval for submission.}

\appendix

\section{Angular momentum for compressible fluids}
\label{appendix:Lcompressible}
We investigate whether alternative definition (\ref{eq:projectionL}), which has proven useful for incompressible flows, can be extended to compressible flows with a spatially varying density $\rho(\boldsymbol{r})$. 
For mathematical tractability, we assume that the density does not vanish on the ellipsoidal boundary.
Then, we expand the velocity of compressible flows using the weighted Helmholtz decomposition in rigid ellipsoids as \citep[e.g.][]{vidal2020acoustic}
\begin{subequations}
\label{eq:weightedHelmholtz}
\begin{equation}
    \boldsymbol{v} = (1/\rho ) \, \nabla \times \boldsymbol{A} + \nabla \Phi, \quad \left . \boldsymbol{v} \boldsymbol{\cdot} \boldsymbol{1}_n \right |_S = 0,
    \tag{\theequation \emph{a,b}}
\end{equation}
\end{subequations}
where $\boldsymbol{A}$ is a vector potential and $\Phi$ is a scalar potential. 
The first subspace represents anelastic flows satisfying $\nabla \boldsymbol{\cdot} (\rho \boldsymbol{v}) = 0$ \citep[e.g.][]{jones2011anelastic}, whereas the irrotational subspace  represents compressible flows with $\nabla \boldsymbol{\cdot} (\rho \boldsymbol{v}) \neq 0$ \citep[such as the acoustic modes without rotation, e.g.][]{vidal2021acoustic}. 
This spectral decomposition has the great advantage of being compatible with the natural inner product of the fully compressible (and anelastic) problem \citep[e.g.][]{sobouti1981potentials,clausen2014elliptical}
\begin{equation}
    \langle \boldsymbol{a}, \boldsymbol{b} \rangle_V = \int_V \rho \boldsymbol{a}^\dagger \boldsymbol{\cdot} \boldsymbol{b} \, \mathrm{d} V,
    \label{eq:InnerProdComp}
\end{equation}
where $\boldsymbol{a}^\dagger$ is the complex conjugate of the vector $\boldsymbol{a}$, contrary to the usual Helmholtz decomposition $\boldsymbol{v} = \nabla \times \boldsymbol{A} + \nabla \Phi$. 
Consequently, the two subspaces in decomposition (\ref{eq:weightedHelmholtz}) are mutually orthogonal with respect to inner product (\ref{eq:InnerProdComp}). 
Guided by planetary applications, we only consider in the following density profiles of the form
\begin{equation}
    \rho(\boldsymbol{r}) = \rho_0(x^2/a^2 + y^2/b^2 + z^2/c^2),
    \label{eq:densityEllip}
\end{equation}
for which the density is constant on every homothetic ellipsoidal shell in the interior. 
Such density profiles are indeed often assumed in compressible planetary models, where they represent background density profiles \citep[e.g. in ellipsoids][]{clausen2014elliptical,vidal2020acoustic}. 

\subsection{Direct calculation}
The angular momentum is defined for compressible fluids as $\boldsymbol{L} = \int_V \boldsymbol{r} \times (\rho_0 \boldsymbol{v}) \, \mathrm{d} V$.
As in the incompressible case, the anelastic subspace has elements with non-zero angular momentum \citep[e.g. in spheres][]{jones2011anelastic}.
Hence, it only remains to calculate the angular momentum associated with the compressible subspace in decomposition (\ref{eq:weightedHelmholtz}). 
A direct calculation gives \citep[using formula B26 in][]{mathews1991forced}
\begin{subequations}
\begin{align}
    \int_V \boldsymbol{r} \times ( \rho_0 \nabla \Phi) \, \mathrm{d} V &= - \int_V \Phi \, (\boldsymbol{r} \times \nabla \rho_0) \, \mathrm{d} V - \int_V \nabla \times \left ( \rho_0 \Phi \, \boldsymbol{r} \right ) \, \mathrm{d}V, \\
    &= - \int_V \Phi \, (\boldsymbol{r} \times \nabla \rho_0) \, \mathrm{d} V  + \int_S \rho_0 \Phi \, (\boldsymbol{r} \times \boldsymbol{1}_n) \, \mathrm{d} S.
\end{align}
\end{subequations}
It shows that, if the density is of the form (\ref{eq:densityEllip}), the compressible subspace has no angular momentum in spheres (since $\nabla \rho_0 \propto \boldsymbol{r}$). 
On the contrary, the compressible subspace in spectral decomposition (\ref{eq:weightedHelmholtz}) has always a non-zero angular momentum in ellipsoids. 

\subsection{Projection approach}
We have outlined that the two subspaces in decomposition (\ref{eq:weightedHelmholtz}) have a non-zero angular momentum in compressible ellipsoids. 
The remaining question is whether, as for incompressible flows, this angular momentum is solely carried by the uniform-vorticity elements $\boldsymbol{e}_i (\boldsymbol{r})$ given by formula (\ref{eq:GP1}) in rigid ellipsoids.
We project the velocity onto the three uniform-vorticity elements with respect to inner product (\ref{eq:InnerProdComp}), obtaining 
\begin{equation}
    \int_V  \boldsymbol{e}_i \boldsymbol{\cdot} (\rho_0 \boldsymbol{v}) \, \mathrm{d} V = \underbrace{\int_V (\boldsymbol{1}_i \times \boldsymbol{r} ) \boldsymbol{\cdot} \rho_0 \boldsymbol{v} \, \mathrm{d} V}_{\boldsymbol{L} \boldsymbol{\cdot} \boldsymbol{1}_i} + \int_V \nabla \Psi_i \boldsymbol{\cdot} (\rho_0 \boldsymbol{v}) \, \mathrm{d} V
    \label{eq:Lcomp1}
\end{equation}
where $\boldsymbol{L} \boldsymbol{\cdot} \boldsymbol{1}_i$ are the Cartesian components of the angular momentum. 
We recover from the above expression that the compressible angular momentum is the projection onto the solid-body rotations $\boldsymbol{1}_i \times \boldsymbol{r}$ in spherical geometries (for which $\Psi_i = 0$). 
An admissible decomposition for compressible spherical flows is thus \citep[e.g.][]{mathews1991forced}
\begin{subequations}
\label{eq:expansioncompBuffett}
\begin{equation}
    \boldsymbol{v}(\boldsymbol{r},t) = \boldsymbol{\omega}(t) \times \boldsymbol{r} + \boldsymbol{v}_f (\boldsymbol{r},t), \quad \int_V \boldsymbol{r} \times (\rho_0 \boldsymbol{v}_f) \, \mathrm{d} V = \boldsymbol{0},
    \tag{\theequation \emph{a,b}}
\end{equation}
\end{subequations}
where the compressible flow $\boldsymbol{v}_f$ has no angular momentum by definition since $\langle \boldsymbol{\omega} \times \boldsymbol{r}, \boldsymbol{v}_f \rangle=0$. 
In ellipsoids, the last volume integral in equation (\ref{eq:Lcomp1}) can be simplified by using the divergence theorem and decomposition (\ref{eq:weightedHelmholtz}).
It gives
\begin{subequations}
\label{eq:Lcomp2}
\begin{align}
    \int_V \nabla \Psi_i \boldsymbol{\cdot} (\rho_0 \boldsymbol{v}) \, \mathrm{d} V &= \underbrace{\int_S \Psi_x \, (\rho_0 \boldsymbol{v}) \boldsymbol{\cdot} \boldsymbol{1}_n \, \mathrm{d} S}_{0} - \int_V \Psi_i \, \nabla \boldsymbol{\cdot} (\rho_0 \boldsymbol{v}) \, \mathrm{d} V, \\ 
    &= \begin{cases}
    0 & \text{if} \quad \nabla \boldsymbol{\cdot} (\rho_0 \boldsymbol{v}) = 0, \\
    - \int_V \Psi_i \, \nabla \boldsymbol{\cdot} (\rho_0 \nabla \Phi) \, \mathrm{d} V & \text{if} \quad \nabla \boldsymbol{\cdot} (\rho_0 \boldsymbol{v}) \neq 0. \\
    \end{cases}
\end{align}
\end{subequations}
Equation (\ref{eq:Lcomp2}) shows that the angular momentum of anelastic flows with $\nabla \boldsymbol{\cdot} (\rho_0 \boldsymbol{v}) = 0$ is rigorously given by $\boldsymbol{L} \boldsymbol{\cdot} \boldsymbol{1}_i = \langle \boldsymbol{e}_i, \boldsymbol{v} \rangle$, as in the incompressible case. 
We can thus extend formula (\ref{eq:expansionBuffett}) to anelastic flows as
\begin{subequations}
\label{eq:expansionanelasticL0}
\begin{equation}
    \boldsymbol{v}(\boldsymbol{r},t) = \boldsymbol{U} (\boldsymbol{r},t) + \boldsymbol{v}_f (\boldsymbol{r},t), \quad \nabla \boldsymbol{\cdot} (\rho_0 \boldsymbol{v}_f ) = 0, \quad \int_V \boldsymbol{r} \times (\rho_0 \boldsymbol{v}_f) \, \mathrm{d} V = \boldsymbol{0},
    \tag{\theequation \emph{a--c}}
\end{equation}
\end{subequations}
where $\boldsymbol{U} (\boldsymbol{r},t)$ is the uniform-vorticity flow given by expression (\ref{eq:uGP1}), and $\boldsymbol{v}_f$ is an anelastic flow with $\langle \boldsymbol{U}, \boldsymbol{v}_f \rangle = 0$ by definition.  
However, in the fully compressible case, the angular momentum cannot be obtained as the projections of the compressible flow onto the uniform-vorticity elements in ellipsoids (because (\ref{eq:Lcomp2}) does not vanish). 
Moreover, we have by virtue of the divergence theorem
\begin{equation}
    \int_V \boldsymbol{e}_i \boldsymbol{\cdot} (\rho_0 \nabla \Phi) \, \mathrm{d} V = - \int_V \Phi \, \nabla \boldsymbol{\cdot} (\rho_0 \boldsymbol{e}_i) \, \mathrm{d} V = -\int_V \phi \, ( \boldsymbol{e}_i \boldsymbol{\cdot} \nabla \rho_0 ) \, \mathrm{d} V = 0
    \label{eq:A9}
\end{equation}
if the density is of the form (\ref{eq:densityEllip}) because $\boldsymbol{e}_i \boldsymbol{\cdot} \nabla \rho_0 \propto \boldsymbol{e}_i \boldsymbol{\cdot} \boldsymbol{1}_n = 0$ on every homothetic ellipsoidal shell in the volume (i.e. not only on the outer ellipsoidal boundary). 
Thus, the compressible subspace can have a non-zero angular momentum that is not carried by the uniform-vorticity elements in ellipsoids (since we have simultaneously $\langle \boldsymbol{e}_i, \nabla \Phi\rangle=0$ and $\int_V \boldsymbol{r} \times (\rho_0 \nabla \Phi) \, \mathrm{d} V \neq \boldsymbol{0}$). 
In such configurations, a possible generalisation of anelastic expansion (\ref{eq:expansionanelasticL0}) to the compressible case could be 
\begin{subequations}
\label{eq:expansioncompL0}
\begin{equation}
    \boldsymbol{v}(\boldsymbol{r},t) = \boldsymbol{U} (\boldsymbol{r},t) + \boldsymbol{v}_f (\boldsymbol{r},t) + \nabla \Phi, \quad \nabla \boldsymbol{\cdot} (\rho_0 \boldsymbol{v}_f ) = 0, \quad \int_V \boldsymbol{r} \times (\rho_0 \boldsymbol{v}_f) \, \mathrm{d} V = \boldsymbol{0},
    \tag{\theequation \emph{a--c}}
\end{equation}
\end{subequations}
where $\boldsymbol{U} (\boldsymbol{r},t)$ is a uniform-vorticity flow given by expression (\ref{eq:uGP1}) in rigid ellipsoids, $\boldsymbol{v}_f$ is an anelastic flow having no angular momentum (i.e. $\rho_0 \boldsymbol{v}_f = \nabla \times \boldsymbol{A}$ but with $\langle \boldsymbol{U}, \boldsymbol{v}_f \rangle = 0$), and $\nabla \Phi$ is a potential flow carrying a non-zero angular momentum even if $\langle \boldsymbol{U}, \nabla \Phi \rangle = 0$ according to equation (\ref{eq:A9}). 

The anelastic and fully compressible cases may thus give different results for the evolution of angular momentum in rotating compressible ellipsoids.
Differences between the two formulations can be expected when the compressible subspace significantly interacts with the anelastic one in spectral decomposition (\ref{eq:weightedHelmholtz}). 
This for instance happens in the presence of global rotation when $M_\Omega = \mathcal{O}(10^{-1})$, where $M_\Omega = R \Omega_0/C_0$ is the rotational Mach number \citep{vidal2020acoustic,vidal2021acoustic} and $C_0$ is the speed of sound. 
Planetary estimates give $M_\Omega = \mathcal{O}(10^{-3})$ for planetary moons, but larger values $M_\Omega = \mathcal{O}(10^{-1})$ are obtained in Jupiter-like gaseous planets
\citep[which are also non-spherical because of centrifugal gravity, e.g.][]{zhang2017shape}. 
Investigating the long-term evolution of angular momentum in such strongly compressible rotating bodies certainly deserves further work.

\section{Viscous decay rates}
\label{appendix:viscous}
\begin{figure}
    \centering
    \begin{tabular}{cc}
    \begin{tabular}{c}
        \includegraphics[width=0.49\textwidth]{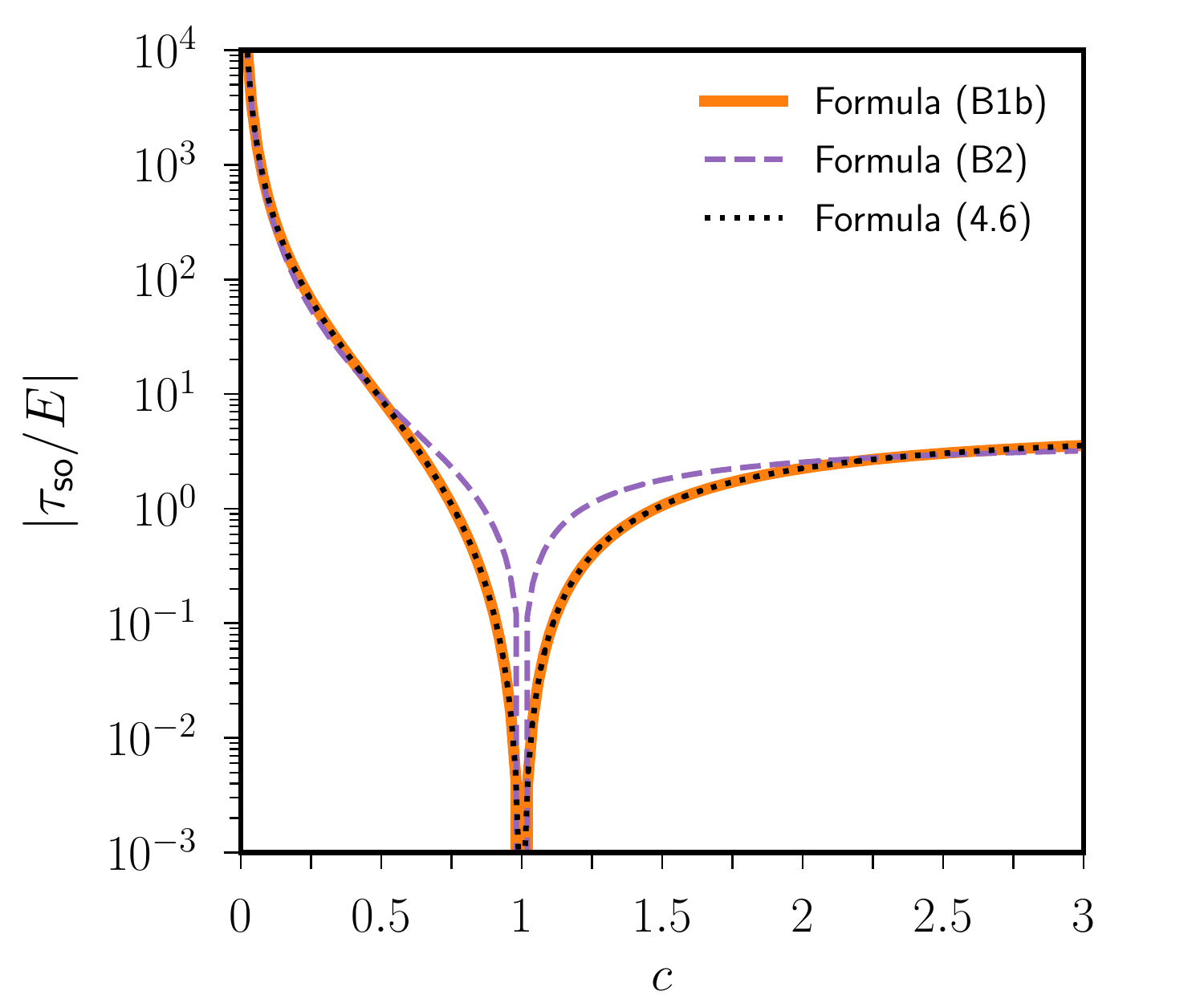}\\
    \end{tabular} &
    \begin{tabular}{c}
        \includegraphics[width=0.49\textwidth]{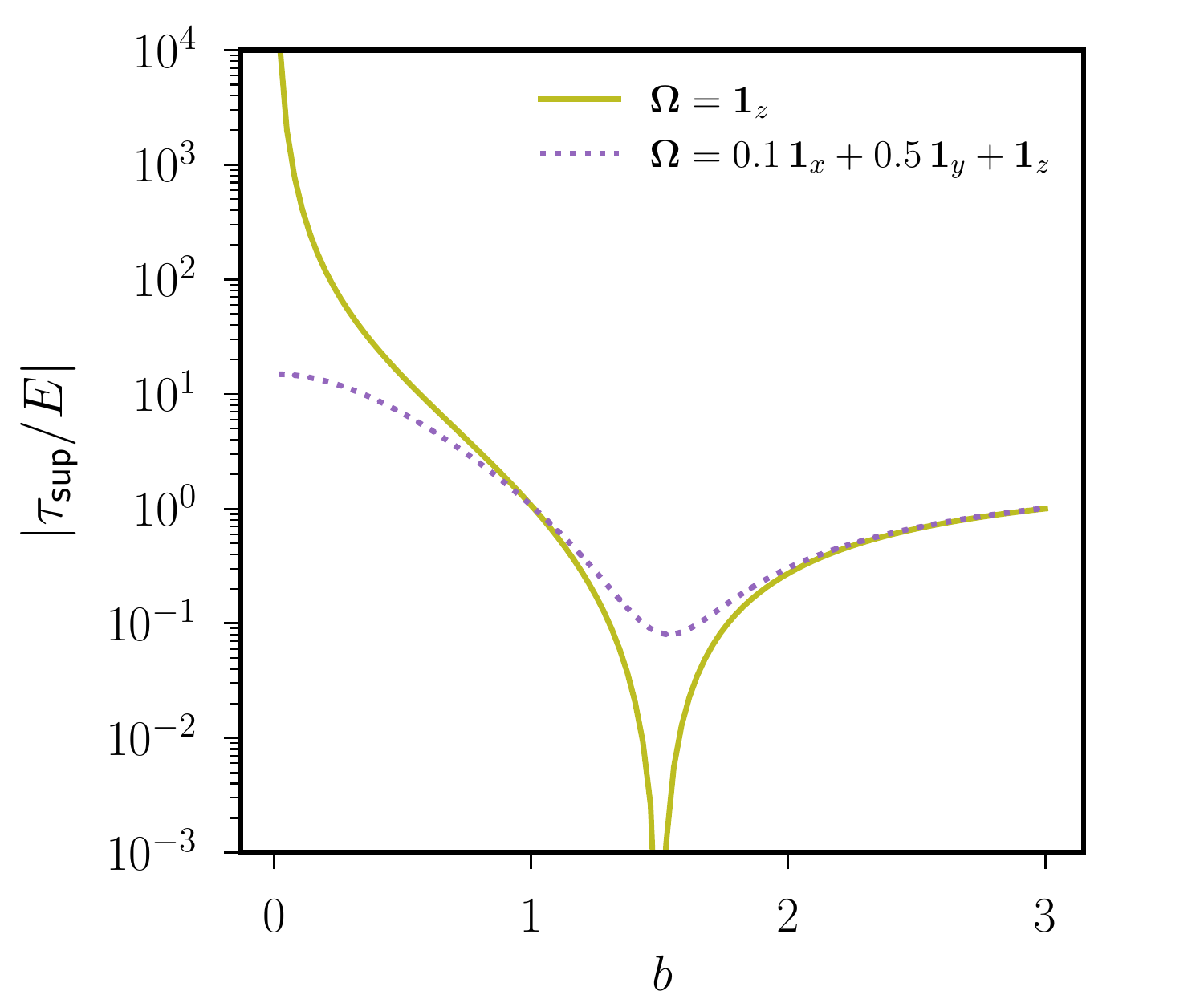}\\
    \end{tabular} \\
    (a) & (b) \\
    \end{tabular}
    \caption{(a) Decay rate $|\tau_\text{so}/E|$ for $\boldsymbol{\Omega}=\boldsymbol{1}_z$ as a function of the semi-axis $c$, in spheroids with $a=b=1$. 
   Comparison between correct formula (\ref{eq:Is}) and erroneous one (\ref{eq:ISLiao2001}) for the surface integral in expression (\ref{eq:decaySF2}). Note that $|\tau_\text{so}| \to 0$ when $c \to 1$. 
   (b) Decay rate $|\tau_\text{sup}/E|$ computed from formula (\ref{eq:decaySFVol}) as a function of the semi-axis $b$, for two rotation vectors $\boldsymbol{\Omega}$ in ellipsoids with $a=1.5$ and $c=1$. Note that $|\tau_\text{sup}| \to 0$ when $b \to a$ (i.e. in the spheroid).}
    \label{fig:fig4}
\end{figure}

We present an alternative formula for the viscous decay rate of the Coriolis eigenmodes in stress-free ellipsoids, which is equivalent to formula (\ref{eq:decaySFVol}). 
To enforce SF-BC (\ref{eq:BCSF}) in equation (\ref{eq:tauSF1}), we employ the curvilinear orthogonal coordinates $[q_1,q_2,q_3]$ (such that the boundary is given by a constant value of $q_1$).
Then, the volume integral can be rewritten using the divergence theorem as
\begin{subequations}
\begin{equation}
        \frac{\tau_k}{E} \int_V |\boldsymbol{Q}_k|^2 \, \mathrm{d} V = I_S -  \int_V |\nabla \times \boldsymbol{Q}_k|^2 \, \mathrm{d}V
        \label{eq:decaySF2}
\end{equation}
with the surface integral ($\mathrm{d}S = h_2 h_3 \, \mathrm{d} q_2 \mathrm{d} q_3$ being the surface element)
\begin{equation}
    I_S = 2 \int_S \left ( \frac{1}{h_1 h_2} \frac{\partial h_2}{\partial q_1} \, |\boldsymbol{Q}_k \boldsymbol{\cdot} \boldsymbol{1}_{q_2}|^2 + \frac{1}{h_1 h_3} \frac{\partial h_3}{ \partial q_1} \, |\boldsymbol{Q}_k \boldsymbol{\cdot} \boldsymbol{1}_{q_3}|^2 \right ) \mathrm{d}S,
    \label{eq:Is}
\end{equation}
\end{subequations}
where $[h_1,h_2,h_3]$ are the curvilinear scale factors and $[\boldsymbol{1}_{q_1}, \boldsymbol{1}_{q_2}, \boldsymbol{1}_{q_3}]$ are the orthogonal basis vectors.
In the sphere, expression (\ref{eq:Is}) reduces to
\begin{equation}
    I_S = 2 \int_S |\boldsymbol{Q}_k \times \boldsymbol{1}_{n}|^2 \, \mathrm{d}S, 
    \label{eq:ISLiao2001}
\end{equation}
recovering formula (3.14) of \citet{liao2001viscous} in the sphere. 
Note that vector expression (\ref{eq:ISLiao2001}) has been erroneously employed in the spheroid \citep[see formula (6.21) in][which is incorrect because of the missing curvature terms]{maffei2017characterization}. 
Expression (\ref{eq:decaySF2}) is very difficult to implement in practice (because of the curvilinear coordinates), contrary to formula (\ref{eq:decaySFVol}) in which the volume integral can be performed fully analytically in ellipsoids \citep[e.g. see formula 50 in][]{lebovitz1989stability}.

For a numerical (cross-validation) benchmark of formulas (\ref{eq:decaySFVol}) and (\ref{eq:decaySF2}), we can compute the decay rate $\boldsymbol{Q}_\text{so}$ of the spin-over mode in spheroidal geometries (i.e. with $a=b=1$).
To do so, we take the formula (3.25) in \citet{vantieghem2014inertial}, giving $\boldsymbol{Q}_\text{so}$ for $\boldsymbol{\Omega} = \boldsymbol{1}_z$ in triaxial ellipsoids, and express it using the curvilinear spheroidal coordinates \citep[e.g. equation 3.1 in][]{cebron2021mean}
\begin{subequations}
\begin{equation}
    x = \eta \mathcal{T} \sin(q_2) \cos(q_3), \quad y = \eta \mathcal{T}  \sin(q_2) \sin(q_3), \quad z = \eta \, (\mathrm{d}_{q_1} \mathcal{T} ) \cos(q_2),
    \tag{\theequation \emph{a--c}}
\end{equation}
\end{subequations}
with $\eta = |1-(c/a)^2|^{1/2}$ and $\mathcal{T} = \cosh(q_1)$ for oblate spheroids (i.e. $a \geq c$) or $\mathcal{T} = \sinh(q_1)$ for prolate spheroids (i.e. $a \leq c$). 
The scale factors are then $h_1=h_2 = \eta [\sinh^2(q_1) + \cos^2(q_2)]^{1/2}$ when $a \geq c$ or $h_1=h_2 = \eta [\cosh^2(q_1) - \cos^2(q_2)]^{1/2}$ when $a \leq c$, and $h_3 = \eta \mathcal{T} \sin(q_2)$. 
The differences between formulas (\ref{eq:Is}) and (\ref{eq:ISLiao2001}) are illustrated in figure \ref{fig:fig4}(a). 
For the particular geometry $a=b=1$ and $c=0.9$, we have $\int_V |\boldsymbol{Q}_k|^2 \, \mathrm{d} V~\simeq~3.36965$, $\int_V |\nabla \times \boldsymbol{Q}_k|^2 \, \mathrm{d}V \simeq 37.64855$ and $I_S \simeq 37.23369$ from formula (\ref{eq:Is}). 
Formulas (\ref{eq:decaySFVol}) and (\ref{eq:decaySF2}) then both predict that $\tau_\text{so}/E \simeq -0.12312$ in this spheroidal geometry (as observed in the figure).  
On the contrary, we would get $I_S \simeq 35.30153$ with formula (\ref{eq:ISLiao2001}), yielding the erroneous value $\tau_\text{so}/E \simeq -0.69652$.
Finally, we show in figure \ref{fig:fig4}(b) the decay rate $\tau_\text{sup}$ for different orientations of the mean rotation axis in triaxial ellipsoids.

\section{Planetary extrapolation for dynamo action}
\label{appendix:planets}
\begin{table}
    \centering
    \begin{tabular}{ccccccc}
         Body & $E$ & $f=\eta/2$ & $\alpha$ [$^\circ$] & $Po$ & $P_x$ & $\epsilon$ \vspace{0.5em} \\
         Earth & $10^{-15}$ & $2.5 \times 10^{-3}$ & $23.5$ & \multirow{2}{*}{$-1.1 \times 10^{-7}$} & $-4.4 \times 10^{-8}$ & $1.7 \times 10^{-5}$ \vspace{0.5em} \\
         Early Earth & $5 \times 10^{-16}$ & $1.0 \times 10^{-2}$ & $17.5$ & {} & $-3.3 \times 10^{-8}$ & $2.5 \times 10^{-6}$  \vspace{0.5em} \\
         \multirow{2}{*}{Moon} & \multirow{2}{*}{$10^{-12}$} & $2.5 \times 10^{-5}$ & \multirow{2}{*}{$1.54$} & \multirow{2}{*}{$-4.0 \times 10^{-3}$} & \multirow{2}{*}{$-2.2 \times 10^{-4}$} & $2.7 \times 10^{-2}$ \vspace{0.5em}  \\
        {} & {} & $3.0 \times 10^{-4}$ & {} & {} & {} & $3.0 \times 10^{-2}$ \vspace{0.5em} \\
        \multirow{2}{*}{Early Moon} & \multirow{2}{*}{$5 \times 10^{-13}$} & $1.0 \times 10^{-4}$ & \multirow{2}{*}{$33.2$} & \multirow{2}{*}{$-3.7 \times 10^{-4}$} & \multirow{2}{*}{$-2.0 \times 10^{-4}$} & $5.5 \times 10^{-1}$ \vspace{0.5em} \\
        {} & {} & $1.2 \times 10^{-3}$ & {} & {} & {} & $6.4 \times 10^{-1}$ \vspace{0.5em} \\
    \end{tabular}
    \caption{Precession forcing in the liquid core of the Earth and Moon. Ekman number $E$ based on the typical viscosity value $\nu=10^{-6} $~m$^2$.s$^{-1}$, polar flattening $f=(a-c)/a$, precession angle $\alpha$. Currently, $f$ is well enough known for the Earth \citep{mathews2002modeling}, but the lunar values of $f$ vary from $f=2.5 \times 10^{-5}$ for a purely hydrostatic Moon \citep{le2011impact} to $f=3.0 \times 10^{-4}$ when considering the present-day non-hydrostatic lithosphere and a liquid core of radius $350$~km \citep{viswanathan2019observational}. 
    Parameters for the Early Moon and Earth, estimated $\sim 4$ Gy ago, are deduced from the current values by considering a spin rate $\Omega_0$ two times larger, leading to values of $E$ twice smaller and of $f$ fourth time larger than the present estimates (due to the centrifugal acceleration in $\Omega_0^2$). 
    Typical estimates for the Moon's history from \citet{cebron2019precessing} and the orbital evolution model of \citet{touma1994evolution}, and for the Early Earth from the low-obliquity scenario in \citet{landeau2022sustaining}.
    }
    \label{tab:planets}
\end{table}

We can crudely estimate the dynamo capability of precession-driven flows using energetic arguments.
To do so, we compute a magnetic Reynolds number $Rm$ as
\begin{subequations}
\label{eq:RmEm}
\begin{equation}
    Rm = \mathcal{U}_\text{topo}/Em, \quad Em = \nu_m/(\Omega_0 R^2),
    \tag{\theequation \emph{a,b}}
\end{equation}
\end{subequations}
where $Em$ is the magnetic Ekman number and $\nu_m \sim 0.5 - 4$~m$^2$.s$^{-1}$ is the magnetic diffusivity of the fluid at typical core conditions \citep[estimated from measurements and computations of the electrical conductivity, e.g. see figure 1 in][]{ohta2021thermal}. 
A necessary condition for large-scale dynamo action is that $Rm \geq \mathcal{O}(10^2)$ in spheres or ellipsoids \citep[e.g.][]{chen2018optimal,holdenried2019trio,vidal2021kinematic}. 
Estimating the magnetic Reynolds number thus crucially depends on the scaling law for the flow strength $\mathcal{U}_\text{topo}$, whose order of magnitude is expected to be given by formula (\ref{eq:Toposat}). 
To be more quantitative, we rewrite formula (\ref{eq:Toposat}) using asymptotic flow (\ref{eq:PrecTriax1}) in the planetary regime $|P_x|\ll 1$, which gives at the leading order in $\eta \ll 1$
\begin{equation}
    \mathcal{U}_\text{topo} \simeq K \epsilon \eta \sim K \begin{cases}
    2 |Po| & \text{when} \quad \alpha = \pi/2, \\
    |\tan(\alpha)| \, \eta & \text{when} \quad \alpha \neq \pi/2, \\
    \end{cases}
    \label{eq:Toposat2}
\end{equation}
where $\alpha$ is the precession angle measured from $\boldsymbol{1}_z$, and $K$ is an unknown numerical prefactor that must be determined for planetary extrapolation.   
We recover from our asymptotic solution that the quantity $\epsilon \eta$ is actually independent of $\eta$ at the leading order when $\alpha=\pi/2$ \citep[e.g. see formula 9.b in][]{horimoto2020conical} and that, when $\alpha \neq \pi/2$, the differential rotation $\epsilon$ becomes independent of $Po$ in the regime $|P_x| \ll 1$ \citep[e.g.][]{williams2001lunar,cebron2019precessing}. 
Moreover, local DNS in periodic shearing boxes, performed at $\alpha=\pi/2$, are actually consistent with the scaling law $\mathcal{U}_\text{topo} \propto 0.1 |Po|$ \citep[see figure 7 in][]{barker2016turbulence}, which is of the form (\ref{eq:Toposat2}) with the numerical constant $K \simeq 0.05$. 
Assume that $K$ is a constant (without further numerical results), we can crudely estimate the dynamo capability of the flows driven by the (topographic) inertial instabilities for realistic planetary conditions by combining equations (\ref{eq:RmEm}) and (\ref{eq:Toposat2}).

Using acceptable scenarios for the lunisolar precession over time (see table \ref{tab:planets}), we obtain $Rm \leq \mathcal{O}(10)$ in the Earth's core over geological ages, showing that precession was not strong enough to drive dynamo action \citep[even billion years ago, which agrees with the conclusions of][]{landeau2022sustaining}. 
Similarly, the estimate $Rm \leq \mathcal{O}(1)$ in the current Moon's core shows precession is not presently dynamo capable \citep[in agreement with the end of the lunar dynamo observed in paleomagnetic studies, e.g.][]{mighani2020end}. 
However, we can obtain larger values $Rm \leq 140$ for the liquid core of the early Moon (depending on the uncertainties on the polar flattening $\eta$ and the magnetic diffusivity). 
Our estimate thus suggests that precession might have been dynamo capable in the early Moon \citep[as initially suggested by][]{dwyer2011long}. 
Further work is obviously needed to rigorously assess the relevance of scaling law (\ref{eq:Toposat2}) in precessing ellipsoids, which is key for planetary extrapolation. 
Adopting SF-BC would be particularly useful to strongly weaken the viscous turbulent flows \citep[which are a priori not well suited to sustain large-scale dynamo fields, see][]{cebron2019precessing} and extract a robust scaling law for the saturation amplitude of the inertial (topographic) instabilities.

\newpage
{
\bibliography{./bib_prec}
\bibliographystyle{jfm}
}

\end{document}